\documentclass[11pt]{article}
\usepackage{latexsym} %\usepackage{epsf}
\usepackage{amsfonts}
\usepackage{amsmath}
\usepackage{amsthm}
\usepackage{amssymb}
\usepackage{url}
\usepackage{hyperref}
\usepackage{color} % prida moznost barevnych textu
\usepackage{stackrel} % pridava moznosti \stackrel, \stackbin
\usepackage[makeroom]{cancel}
\renewcommand{\theequation}{\thesection.\arabic{equation}}
\newcounter{subequation}[equation]
\makeatletter

\newcommand{\p}{^{\prime}}

\expandafter\let\expandafter\reset@font\csname reset@font\endcsname

\def\subeqnarray{\arraycolsep1pt
    \def\@eqnnum\stepcounter##1{\stepcounter{subequation}%
        {\reset@font\rm(\theequation\alph{subequation})}}
\jot5mm     \eqnarray}

\makeatother

\def\u{\hbox{\rm u}}

\def\be{\begin{equation}}
\def\ee{\end{equation}}
\def\bea{\begin{eqnarray}}
\def\eea{\end{eqnarray}}
\def\ba{\begin{array}}
\def\ea{\end{array}}
\def\dd{\partial}

\def\eps{\varepsilon}
\def\half{\frac{1}{2}}

\def\one#1{#1^{\raise5pt\hbox{$\scriptstyle\!\!\!\!1$}}\,{}}
\def\two#1{#1^{\raise5pt\hbox{$\scriptstyle\!\!\!\!2$}}\,{}}

\def\tilde{\widetilde}
\def\II{\hbox{{1}\kern-.25em\hbox{l}}}
\makeatletter
\def\binrel@#1{\begingroup
  \setboxz@h{\thinmuskip0mu
    \medmuskip\m@ne mu\thickmuskip\@ne mu
    \setbox\tw@\hbox{$#1\m@th$}\kern-\wd\tw@
    ${}#1{}\m@th$}%
  \edef\@tempa{\endgroup\let\noexpand\binrel@@
    \ifdim\wdz@<\z@ \mathbin
    \else\ifdim\wdz@>\z@ \mathrel
    \else \relax\fi\fi}%
  \@tempa
}
\let\binrel@@\relax
\def\overset#1#2{\binrel@{#2}%
  \binrel@@{\mathop{\kern\z@#2}\limits^{#1}}}
\def\underset#1#2{\binrel@{#2}%
  \binrel@@{\mathop{\kern\z@#2}\limits_{#1}}}
\makeatother
\newfont{\bbd}{msbm10 scaled\magstep1}

\def\R{\hbox{\bbd R}}

\def\e{\varepsilon}
\begin{document}

\begin{center}
{\LARGE {Correlators with $s\ell_2$ Yangian symmetry 
}}  

 \vspace{0.3cm}

\large \sf J. Fuksa$^{a,c,}$\footnote{\sc e-mail: fuksa@theor.jinr.ru}, \
R. Kirschner$^{b,}$\footnote{\sc e-mail: Roland.Kirschner@itp.uni-leipzig.de} \\

\vspace{0.5cm}

\begin{itemize}
\item[$^a$]
{\it Bogoliubov Laboratory of Theoretical Physics,\\ Joint Institute for Nuclear Research, Dubna, Russia}
\item[$^b$]
{\it Institut f\"ur Theoretische 
Physik, Universit\"at Leipzig, \\
PF 100 920, D-04009 Leipzig, Germany}
\item[$^c$]
{\it Faculty of Nuclear Sciences and Physical Engineering,\\ Czech Technical University in Prague, Czech republic}
\end{itemize}
\end{center} 

\vspace{2 cm}   
\begin{abstract}
\noindent
Correlators based on $s\ell_2$ Yangian symmetry and its
quantum deformation are studied. 
Symmetric integral operators can be defined with such
correlators as kernels. Yang-Baxter operators can be represented in this
way. Particular Yangian symmetric correlators are related to the kernels of 
QCD parton evolution. The solution of the eigenvalue problem of Yangian
symmetric operators is described. 
\end{abstract}

\newpage

%%%%%%%%%%%%%%%%%%%%%%%%%%%%%%%%%%%%%%%%%%%%%%%%%%%%%%%%%%%%%%%%%%%%%%%%%%%%%%
%{\small \tableofcontents}
\renewcommand{\refname}{References.}
\renewcommand{\thefootnote}{\arabic{footnote}}
\setcounter{footnote}{0}

\section{Introduction}
The symmetry on which quantum integrable systems (QIS) like spin chains are based 
can be formulated in terms of a Yangian algebra \cite{Drinfeld}, which
results from the expansion of the  monodromy matrix.
 In the applications of QIS to gauge field theories, which
attracted much attention in the last two decades, the symmetry appears
typically in infinite-dimensional representations with action on functions. 
The first of such applications concerned the high-energy asymptotics of
scattering in Quantum Chromodynamics and are based on $s\ell_2$
\cite{LevPadua,BDM}. In the
Regge asymptotics \cite{BFKL} the dynamics reduces to the plane transverse to the
scattering axis and in the Bjorken asymptotics \cite{DGLAP,ERBL} 
the dynamics reduces
to a light ray. The latter asymptotics is connected with 
the scale dependence of composite operators. 
 In the case of $\mathcal{N}=4$ supersymmetric
Yang-Mills theory the relation to QIS has been shown \cite{BSt} to work with the
super $s\ell(4|4)$ spin chain for all composite operators in all orders 
of perturbative  expansion and
to be connected to strings via the AdS/CFT correspondence, where in the
supergravity on the string side classical integrable systems appear.
The advances in the methods of scattering amplitude computations have lead in
particular for the $\mathcal{N}=4$ supersymmetric
Yang-Mills theory to the discovery of the dual superconformal symmetry \cite{KS}
and the understanding that the two superconformal symmetries are embedded in
the corresponding Yangian algebra \cite{DHP}. 
 
Yangian symmetric correlators (YSC) have been proposed
as a convenient formulation of the Yangian symmetry of scattering
amplitudes \cite{CK}. They are defined by an eigenvalue relation involving the
monodromy matrix operator. A number of properties, previously observed in
amplitude calculations, have been shown to follow from this monodromy
condition. There are convenient methods of construction of YSC, the most important
one is based on Yang-Baxter $RLL$ type relations. The construction methods are
related to the on-shell graph method developed in the amplitude context.
This relation has been  discussed in detail in \cite{Broedel}.

The tool of YSC is not restricted in its application to super Yang-Mills
scattering amplitudes. It may be used wherever Yangian symmetry appears,
 it is powerful and  it will help to exploit the
higher symmetry in more problems. 
The investigation of the particular case of 
 $s\ell_2$ is important for testing the approach, first of all because this
case is relevant in a number of integrable models,
in two-dimensional conformal field theories \cite{BazhanovZam} 
and in particular questions of gauge field theory.   The advantage of the
method should be tested here in comparison with the standard methods,
reproducing known results in alternative ways and going beyond them by
relying on the new viewpoint provided.
The second reason for devoting attention to the $s\ell_2$ case  is 
the fact that here Yang-Baxter relations are known in several versions 
and in most explicit form.

In the present paper we expand the range for Yangian symmetric correlators.
We shall show that the approach of YSC is powerful also beyond the investigation 
of super Yang-Mills scattering  amplitudes, where it has been introduced. 
We shall demonstrate its flexibility by indicating several ways of
calculation,  several  explicit formulations, the opportunity to perform
algebraic deformations. 

In examples for the $s\ell_2$ case we shall
demonstrate new construction methods beyond the ones known in the amplitude
context and the role of the spectral parameters characterizing the
representation of the Yangian algebra.   

The approach allows to treat the Yangian symmetry in 
a convenient way.  We shall discuss the role of YSC as kernels of symmetric
operators. The symmetry of the correlators induces the symmetry of
the integral operators.

The approach has a high potential of physical applications.
We shall show how the Yangian symmetry of the correlators leads to the exact
solution of the spectral problem for the related operators, which is of
great importance in view of physics. 
Among many applications related to the $s\ell_2$ case we pick up the example
of the scale dependence of  parton distributions and composite operators.  
We shall establish a direct relation of the kernels of the leading order
scale evolution to YSC.

In Sect. \ref{sec:2} we recall the definition of YSC and formulations of Yang-Baxter
relations.
We use the explicit expressions of Yang-Baxter (YB) operators together with
methods developed for amplitudes \cite{AH} to obtain explicit expressions
of YSC. We emphasize the specifics of the case of $s\ell_2$, which allows
several versions in the construction in particular those differing
essentially from the ones close to the amplitude methods.  A number of
explicit examples will be given for illustration in Sect. \ref{sec:3}.

We keep arbitrary spectral parameters in the monodromy matrix from
which the YSC construction starts, explain their relation to the dilatation
weight associated with each point and also how they characterize the
representation of the Yangian algebra. Note that in $\mathcal{N} =4$ super Yang-Mills
amplitudes the case of vanishing spectral parameters matters and therefore the case of
arbitrary values has been considered in a few papers only, e.g.
\cite{FLMPS13,Broedel,FLS14,BBR14}.  
We shall see that spectral parameters play an active role in the method.

The symmetry condition defining YSC generalizes to algebraic
deformations. We demonstrate this in the trigonometric case in Sect. \ref{sec:4}. 

In Sect. \ref{sec:5}
we study Yangian symmetric operators defined  as integral operators with
YSC as kernels. We show how particular 4-point YSC result in the Yang-Baxter
operators. Generalized YB operators result from particular $2M$-point
YSC. 

It is known how YB relations allow to derive the spectrum of the
involved YB $R$ operator. In Sect. \ref{sec:6}
we show that this works also for generalized YB operators
by demonstrating it in the simplest non-trivial case with $M=3$.

The application of QIS and  Yangian symmetry to the scale dependence of
composite operators 
and of (generalized) parton distributions \cite{SPD}
is well known in the formulation of
quantum spin chain dynamics. In Sect. \ref{sec:5}
we present the formulation in terms of
YSC. We identify the kernels of the scale evolution of (generalized) parton
distributions as  4-point YSC with particular values of the dilatation
weights and the spectral parameters. 
In the presented form of ratio-coordinates interpreted as positions on the
light ray these kernels describe the scale dependence of  operators composed
of two fields with derivatives in the two-point light ray form. 

The generalized Yang-Baxter operators constructed from $2M$-point YSC
are relevant for the scale evolution of operators composed out of $M$
fields. Their spectra, the calculation of which we show in the particular case
$M=3$, allows to derive exactly the related anomalous dimensions.  
In the situation of a low number of fields
this approach becomes an alternative to the Bethe ansatz calculations used in the
spin chain formulation.

\section{Yangian symmetry } \label{sec:2}
\setcounter{equation}{0}

\subsection{Monodromy matrix operators and symmetric correlators}

We start with the formulation of the $s\ell_2$ Lie algebra relations as the
fundamental Yang-Baxter relation
\be \label{fundYB}
 \mathcal{R}(u-v) (L(u)\otimes I) (I\otimes L(v)) = (I\otimes L(v))( L(u)\otimes I) \mathcal{R}(u-v).
\ee
Here $\mathcal{R}(u-v)$ stands for Yang's $R$ matrix, which is $4\times 4$
in our case and  composed of the unit matrix and the matrix representing the
permutation of the tensor factors in the tensor product of the
two-dimensional fundamental representation spaces,  
$ \mathcal{R}(u) = u I_{4\times 4} + P $. The $L$ matrices are linear in the
spectral parameter $u$, $L(u) = u I_{2\times 2} +  L $. 
With this ansatz 
(\ref{fundYB}) implies that the matrix elements of $ L$ generate the
$s\ell_2$ algebra. We shall introduce several forms of $L$ below. 

The Yangian algebra can be introduced by the matrix operator $T(u)$
obeying the above YB relation with  $L(u)$  substituted by $T(u)$ and where now
the dependence on $u$ is not restricted as above. Then the generators of the
Yangian algebra \cite{Drinfeld} appear in the  expansion of $T(u)$ 
in inverse powers of $u$ and the
Yangian algebra relations  follow from the fundamental YB relation
(\ref{fundYB}).
We shall deal with  evaluations  of the $s\ell_2$ Yangian of finite order $N$
where $T(u)$ is constructed from the $L$ matrices in the way well know in 
the context of integrable models. 
Let the matrix elements of $L_i(u_i)$ act on the representation space $V_i$
and consider  the matrix product defining the monodromy matrix    
$$ T(u,\delta_1, \dots,\delta_N)= T(\mathbf{u}) = L_1(u_1) \cdots L_N(u_N), \quad u_i =
u+\delta_i,  $$ 
acting on the tensor product $ V_1 \otimes \cdots \otimes V_N$. In the case if
the representations $V_i$ are of definite weight $2 \ell_i$
the weight parameters are added to the parameter list abbreviated by $\mathbf{u}$.
We consider  infinite-dimensional representations $V_i$ in terms of
functions of two or one variables. In the case of $V_i$ of definite weight the
functions of two variables $\mathbf{x}_i = (x_{i,1}, x_{i,2}) $ are homogeneous 
and the weight $2\ell_i$ is the degree of homogeneity,
$$ \psi(\mathbf{x}_i) = x_{i,2}^{2\ell_i} \phi  (x_i) = x_{i,1}^{2\ell_i}
\tilde \phi(y_i). $$
The one-dimensional variables are related to the two-dimensional
as ratios of the latter components, where we have two versions in the
following discussion,
$ x_i = \frac{x_{i,1}}{x_{i,2}}$ and $ y_i = - \frac{x_{i,2}}{x_{i,1}} 
$.
The elements of the tensor product $ V_1 \otimes \cdots \otimes V_N$ are then 
functions of $N$ points,  two- or one-dimensional.

$N$-point functions $\Phi $ obeying the condition
\be \label{YSC}
 T(\mathbf{u}) \Phi = I E(\mathbf{u}) \Phi \ee
are called Yangian symmetric correlators (YSC). 
On l.h.s an operator-valued matrix is acting and on r.h.s. there is the unit
matrix $I$ and  the eigenvalue  depending on the set of parameters
$\mathbf{u}$.   
 As solutions we shall accept also expressions involving
distributions, extending the representation space $ V_1 \otimes \cdots \otimes
V_N$.  The $\delta$-distributions are to be understood in Dolbeault sense
(cf. \cite{GH}), their arguments  must not be restricted to real
values.

The eigenvalue $E(\mathbf{u})$ is  related to the quantum determinant 
of the monodromy $T(\mathbf{u})$. In the $s\ell_n$ case the definition is 
$$
\mathop{\mathrm{ qdet}}(T(\mathbf{u})) = 
\sum_{\sigma\in S_n} \mathop{\mathrm{ sgn}}(\sigma) T_{1\sigma(1)}({u}) 
T_{2\sigma(2)}({u-1}) \cdots T_{n\sigma(n)}({u-n+1}),
$$
where the sum is performed over the symmetric group $S_n$ of the order $n$.
On r.h.s. $T_{\alpha \beta}(u-k)$ are the matrix elements of 
$T(u, \delta_1, ...,\delta_N)$. 
The quantum determinant $\mathop{\rm qdet}(T(\mathbf{u}))$ is the generating 
function for the center of  Yangian $Y(g\ell_n)$.
Because the off-diagonal elements of  $T(\mathbf{u})$ annihilate the function 
$\Phi$ it follows immediately from \eqref{YSC} that in our case of $n=2$
\begin{align}
\mathop{\rm qdet}(T(\mathbf{u})) \Phi   = 
T_{11}({u}) T_{22}({u-1})  \Phi \notag 
 = E({u})E({u-1}) \Phi. \notag
\end{align}

\subsection{The homogeneous form}

We start with two forms of the $L$ operators,
$L_i^{\pm}(u) = I u + L_i^{\pm}(0)$, with matrix elements  
\be \label{L+-}
 L^+_{i, \alpha \beta} (0) = \dd_{i,\alpha} x_{i,\beta}, \  \ \  
L^-_{i, \alpha \beta}(0) =  - x_{i, \alpha} \dd_{i,\beta}
\ee
built from two canonical pairs $\mathbf{x}_i, \mathbf{p}_i$ at each point
$i$, $\mathbf{x}_i=(x_{i,1}, x_{i,2}), \mathbf{p}_i = (\dd_{i,1}, \dd_{i,2})$.
Notice that the elementary canonical transformation
\be \label{canon} \begin{pmatrix}
x_{i,\alpha} \\ \dd_{i,\alpha} 
\end{pmatrix} \rightarrow
 \begin{pmatrix}
\dd_{i,\alpha} \\ -x_{i,\alpha} 
\end{pmatrix} \ee
transforms $L_i^+(u)$ into $L_i^-(u)$. Here the transformation acts
uniformly for the index values $\alpha = 1,2$.  With a transformation modified
in this respect, i.e. non-trivial for the index value $\alpha = 1$ only, 
\be \label{canon1}
 \begin{pmatrix}
x_{i,1} \\ \dd_{i,1} 
\end{pmatrix} \rightarrow
 \begin{pmatrix}
\dd^{\lambda}_{i} \\ -\lambda_{i} 
\end{pmatrix},  \ \ \ 
\begin{pmatrix}
x_{i,2} \\ \dd_{i,2} 
\end{pmatrix} \rightarrow
\begin{pmatrix}
\bar \lambda_{i} \\ \bar \dd^{\lambda}_{i} 
\end{pmatrix}, 
\ee
applied to $L^+_i(u)$ we are lead to the helicity form{\footnote
{We use here the term helicity form in analogy to the case of $s\ell_4$
Yangian symmetry, where it has been  applied to scattering
amplitudes.}}
$$ L^{\lambda}_i(0) =
\begin{pmatrix}
- \lambda_{i}\dd^{\lambda}_{i} & -\lambda_{i}\bar \lambda_{i} \\
\dd^{\lambda}_{i}\bar \dd^{\lambda}_{i} & \bar \dd^{\lambda}_{i} \bar \lambda_{i} 
\end{pmatrix}.
$$
We have the matrix inversion relations
\be \label{inv}
 \left ( \frac{L_i^+ (u)}{u} \right )^{-1} = \frac{L_i^+(-u-1-
(\mathbf{x}_i\mathbf{p}_i))}{
-u-1- (\mathbf{x}_i\mathbf{p}_i)},  
\ee
$$
\left ( \frac{L_i^- (u)}{u} \right )^{-1} = \frac{L_i^-(-u+1 +
(\mathbf{x}_i\mathbf{p}_i))}{
-u+1+ (\mathbf{x}_i\mathbf{p}_i)},   
$$
where $(\mathbf{x}_i\mathbf{p}_i) = x_{i,1} \dd_{i,1} + x_{i,2} \dd_{i,2}$. 
We work with representations on functions of $\mathbf{x}_i = ( x_{i,1},
x_{i,2})$ or $\lambda, \bar \lambda $ so that $\mathbf{p}_i= ( \dd_{i,1},
\dd_{i,2}) $ or $ \dd^{\lambda}_{i}, \bar \dd^{\lambda}_{i}$ act as
derivatives as suggested by notation.  

We consider the YB relations of $RLL$ type 
$$ R^{++}_{12}(u_1-u_2) \ L_1^+(u_1)L_2^+(u_2) = L_1^+(u_2)L_2^+(u_1) \
R^{++}_{12}(u_1-u_2),
$$
$$ R^{+-}_{12}(u_1-u_2) \ L_1^+(u_1)L_2^-(u_2) = L_1^+(u_2)L_2^-(u_1) \
R^{+-}_{12}(u_1-u_2),
$$ 
\be \label{YBu} 
R^{\lambda}_{12}(u_1-u_2) \ L_1^{\lambda}(u_1)L_2^{\lambda}(u_2) = 
L_1^{\lambda}(u_2)L_2^{\lambda}(u_1) \
R^{\lambda}_{12}(u_1-u_2),
\ee 
and find the following expressions for the YB operators:
\be \label{R12}
R^{+-}_{12}(u) = (\mathbf{x}_1 \cdot \mathbf{x}_2)^u , \ \ \ 
R^{++}_{12}(u) = \int \frac{dc}{c^{1+u}} 
e^{-c (\mathbf{x}_1 \cdot \mathbf{p}_2)}. \ee
Here the integration contour is arbitrary with the restriction that
integration by parts is performed with vanishing boundary terms (Appendix \ref{app:A}).
The expression for $R^{\lambda}_{12}$ is obtained from the one for
$R^{++}_{12}(u)$ by substituting according to the canonical
transformation (\ref{canon1})
$$ (\mathbf{x}_1 \cdot \mathbf{p}_2) \to 
-\dd^{\lambda}_{1} \lambda_{2} + \bar \lambda_{1} \bar \dd^{\lambda}_{2}.
$$

If the representation spaces $V_1,V_2$ are restricted to the subspaces of 
 eigenvalues $2\ell_i$ of $(\mathbf{x}_i\mathbf{p}_i)$
 then the inversion relations (\ref{inv}) imply further
Yang-Baxter relations. The dependence on the eigenvalue of the restricted 
 $L$ matrices  is conveniently formulated by a second
argument $u^+= u + 2\ell$ or $u^- = u-2\ell-2$ as
$$ L^+(u)|_{2\ell} = L^+(u^+, u), \ \ L^-(u)|_{2\ell} = L^-(u, u^-). $$
The additional YB relations read
\be \label{YBu2}
 R^{+-}_{21}(u_1^--u_2^+) \ L_1^-(u_1, u_1^-)L_2^+(u_2^+,u_2) 
= L_1^-(u_1, u_2^+)L_2^+(u_1^-,u_2) \ R^{+-}_{21}(u_1^--u_2^+),
\ee
$$ R^{++}_{21}(u_1^+-u_2^+) \ L_1^+(u_1^+,u_1)L_2^+(u_2^+,u_2) = 
L_1^+(u_2^+,u_1)L_2^+(u_1^+,u_2) \ R^{++}_{21}(u_1^+-u_2^+).
$$
 The previous relations read in the new form
\be \label{YBu1}
 R^{+-}_{12}(u_1-u_2) \ L_1^+(u_1^+,u_1)L_2^-(u_2, u_2^-) = 
L_1^+(u_1^+,u_2)L_2^-(u_1, u_2^-) \ R^{+-}_{12}(u_1-u_2),
\ee
$$ R^{++}_{12}(u_1-u_2) \ L_1^+(u_1^+, u_1)L_2^+(u_2^+,u_2) = 
L_1^+(u_1^+, u_2)L_2^+(u_2^+,u_1) \ R^{++}_{12}(u_1-u_2).
$$
 Each of the relations is characterized by the exchange of a pair of
parameters.  We shall consider the transformation of monodromy matrices as
$ T(\mathbf{u}) R = R T(\mathbf{u}\p) $
accompanied by a change of the parameters by permutations $\sigma, \bar
\sigma$ of the sets
$ u_1, \dots,u_N$ and $u^+_1, \dots, u_N^+$
 This is conveniently written as the table in two rows
$$ \begin{pmatrix}  
u_{\sigma(1)} & u_{\sigma(2)} & \dots & u_{\sigma(N)} \\
u^+_{\bar \sigma(1)}& u^+_{\bar \sigma(2)} & \dots & \u^+_{\bar \sigma(N)} 
\end{pmatrix}.
$$
We shall use also the abbreviation  by writing the indices carried by the
parameters only.

\subsection{The ratio-coordinate form} \label{sec:2.3}

The reduction of the $L$ operators to the subspace of eigenvalue
$2\ell$  is done  for the  action  on functions of
the coordinate components by the restriction to a definite degree of homogeneity.
$$
 \psi_{2\ell} (\mathbf{x}) =( x_2)^{2\ell}  \phi(x), \ \
L^+(u) ( x_2)^{2\ell} \phi(x) = ( x_2)^{2\ell}   L_x(u^+ +1,u) \phi(x), $$
\be\label{Lred}  \psi_{2\ell} (\mathbf{x}) =( x_1)^{2\ell}  \tilde \phi(y), \ \
L^-(u) ( x_1)^{2\ell} \tilde \phi(y) = ( x_1)^{2\ell}   L_y(u, u^- +1) \tilde
\phi(y),   \ee
$$ x = \frac{x_1}{x_2}, \ y = - \frac{x_2}{x_1}. $$
The conjugated momenta are to be substituted in the  case $+$ as
$$ x_1 \dd_1 \to x \dd, \ 
x_2 \dd_2 = (\mathbf{x}\mathbf{p}) - x_1 \dd_1 \to 2\ell - x \dd,
$$
and in the  case $-$ as 
$$ x_1 \dd_1 = (\mathbf{x}\mathbf{p}) - x_2 \dd_2 \to 2\ell - y \dd, \ 
x_2\dd_2 \to y \dd. $$

Explicitly we have
$$ L^+(u) =
\begin{pmatrix}
  u+1 + x_1 \dd_1 & x_2 \dd_1 \\
x_1 \dd_2 & u+1 + x_2 \dd_2 
\end{pmatrix}, \ $$ $$
  L_x(u^++1, u) =
\begin{pmatrix}
  u+1 + x \dd &  \dd  \\
x (- x \dd + 2\ell)  & u+1 + 2\ell - x \dd 
\end{pmatrix},
$$
$$
L^-(u) =
\begin{pmatrix}
  u - x_1 \dd_1 & - x_1 \dd_2 \\
-x_2 \dd_1 & u - x_2 \dd_2 
\end{pmatrix},
\ \ \  L_y(u,u^-+1) =
\begin{pmatrix}
  u -2\ell + y \dd &   \dd  \\
y (- y \dd + 2\ell)  & u  - y \dd
\end{pmatrix}.
$$
In this way the $L$ operators in the two dual representations of $g\ell_2$ result in 
the same form  of $L$ operators of $s\ell_2$,
$L(u, u^- +1) = L(\tilde u^++1,\tilde u)$ for $ \tilde u = u^- +1 = u-2\ell-1$. 
Therefore we can omit the superscripts $\pm$ in the ratio-coordinate form.
The 
uniformization of both results is achieved also in the notation used in
\cite{DKK07}. We substitute for the signature $+$ case $v= u+\ell $, but for the
signature $-$ case $v= u-\ell-1$. Then the reduced $L$ appears in both cases 
with the same arguments, $L(v^{(1)}+1, v^{(2)}+1)$, where
\be \label{v1v2}
 v^{(1)} = v + \ell, v^{(2)} = v-\ell-1. \ee
The matrix $L(v^{(1)}, v^{(2)})$ factorizes as 
\be \label{Lratio} 
L(v^{(1)},v^{(2)}) =
\begin{pmatrix}
  v^{(2)}+1 + x \dd &  \dd  \\
x (- x \dd + 2\ell)  & v^{(1)} - x \dd 
\end{pmatrix} = \ \  
v^{(1)} \hat V^{-1}(v^{(1)}) \hat D \hat V( v^{(2)}),  \ee
$$ \hat V(v) = 
\begin{pmatrix}
v & 0 \\
-x & -1 
\end{pmatrix}, \ \ 
\hat D = 
\begin{pmatrix}
1 & - \dd \\
0 & 1 
\end{pmatrix}.
$$
This is one argument in favor of the choice of the parameters $v$
with the particular shift with respect to the original ones $u$. The other
(related) argument is the permutation action discussed below.

The $L$ matrix and the $s\ell_2$ generators are related as
$$ L(v^{(1)},v^{(2)}) = 
\begin{pmatrix}
v + S^0 & S^- \\
S^+ & v - S^0 
\end{pmatrix}, 
$$
\be \label{Sa}
 S^- = \dd, \ \ S^+ = x(-x\dd + 2 \ell),  \ \ S^0 = x\dd -\ell. \ee

The inversion relations (\ref{inv}) both read in the notation 
$L(v^{(1)}, v^{(2)})$
\be \label{invv12}
L^{-1}(v^{(1)}, v^{(2)})= - \frac{1}{v^{(1)}v^{(2)}}
L(- v^{(2)},-v^{(1)}) .
\ee
This can also be checked directly starting from the factorized form
(\ref{Lratio}).

The YB relations formulated above (\ref{YBu}), (\ref{YBu2}), (\ref{YBu1}) 
can be rewritten in the ratio coordinates with the notation $L(v^{(1)}, v^{(2)})$ 
which does not distinguish signature.   
The resulting forms of the YB relations are
characterized by particular permutations of the parameters
$v_1^{(1)},v_1^{(2)},v_2^{(1)},v_2^{(2)}$ appearing as arguments of the $L$ operators 
in the product
$L_1(v_1^{(1)},v_1^{(2)}) L_2(v_2^{(1)},v_2^{(2)})$, \cite{SD05}. 
$$  R^1_{12}(v_1^{(1)}|v_2^{(1)}, v_2^{(2)}) 
L_1(v_1^{(1)},v_1^{(2)}) L_2(v_2^{(1)},v_2^{(2)})  =
L_1(v_2^{(1)},v_1^{(2)}) L_2(v_1^{(1)},v_2^{(2)})  R^1_{12}(v_1^{(1)}|v_2^{(1)},
v_2^{(2)}),  $$
$$  R^2_{12}(v_1^{(1)}, v_1^{(2)}| v_2^{(2)}) 
L_1(v_1^{(1)},v_1^{(2)}) L_2(v_2^{(1)},v_2^{(2)})  =
L_1(v_1^{(1)},v_2^{(2)}) L_2(v_2^{(1)},v_1^{(2)})
R^2_{12}(v_1^{(1)}, v_1^{(2)}| v_2^{(2)}). $$
It is appropriate to consider the YB operators related to the elementary
permutations \cite{DKK07}, 
$$  \mathcal{S}_{11} (v_1^{(1)}-v_1^{(2)}) 
L_1(v_1^{(1)},v_1^{(2)}) L_2(v_2^{(1)},v_2^{(2)})  =
L_1(v_1^{(2)},v_1^{(1)}) L_2(v_2^{(1)},v_2^{(2)})
\mathcal{S}_{11} (v_1^{(1)}-v_1^{(2)}), $$
$$  \mathcal{S}_{12} (v_1^{(2)}-v_2^{(1)}) 
L_1(v_1^{(1)},v_1^{(2)}) L_2(v_2^{(1)},v_2^{(2)})  =
L_1(v_1^{(1)},v_2^{(1)}) L_2(v_1^{(2)},v_2^{(2)})
\mathcal{S}_{12} (v_1^{(2)}-v_2^{(1)}),  $$
$$  \mathcal{S}_{22} (v_2^{(1)}-v_2^{(2)}) 
L_1(v_1^{(1)},v_1^{(2)}) L_2(v_2^{(1)},v_2^{(2)})  =
L_1(v_1^{(1)},v_1^{(2)}) L_2(v_2^{(2)},v_2^{(1)})
\mathcal{S}_{22} (v_2^{(1)}-v_2^{(2)}). $$
In the $RLL$ relations the $R$ operators
$R^1$ and $R^2$ resulting from the above $R_{12}^{++}(u_1-u_2)$ and
$R_{21}^{++}(u_1^+ - u_2^+)$ interchange the parameters with superscript 1 or 2
respectively. The YB operator $\mathcal{S}_{12}(v_1^{(2)}- v_2^{(1)}) $
interchanging the adjacent $v_1^{(2)}, v_2^{(1)}$ results from
$R_{12}^{+ -}$ or $R_{21}^{+-}$. Using (\ref{Lratio}) it is obtained as 
\be \label{S12}
\mathcal{S}_{12}(v) = (x_1 - x_2)^{v}. \ee
Notice that here $x_i$ are the ratio coordinates and the subscript is the
point label.
The other elementary permutations
of $v_1^{(1)},v_1^{(2)}$ or of $v_2^{(1)},v_2^{(2)}$ appear in the $RLL$ relations
with $\mathcal{S}_{11}(v_1^{(1)}- v_1^{(2)}) $ or 
$\mathcal{S}_{22}(v_2^{(1)}- v_2^{(2)}) $,
respectively.  They are given by the intertwiner between the representations
of weights $2\ell$ and $-2\ell - 2$.   
$$  \mathcal{S}_{11} (v_1^{(1)}-v_1^{(2)}) = W_1( 2\ell_1+1), \ \ 
\mathcal{S}_{22}(v_2^{(1)}-v_2^{(2)}) = W_2(2\ell_2+1), $$
\be \label{intertW}
 W_i(a) = x_i^{-a} \frac{\Gamma(x_i\dd_i +1) }{\Gamma(x_i\dd_i +1-a)}. \ee
The $R$ operators related to the permutations of parameters of superscript 1
or 2 factorize into the operators $\mathcal{S}_{ij}$
related to the elementary permutations.
$$ R^1_{12}(u^{(1)}|v^{(1)}, v^{(2)}) = \mathcal{S}_{12}(v^{(2)}-v^{(1)})
\mathcal{S}_{22}(u^{(1)}-v^{(1)})   \mathcal{S}_{12}(u^{(1)}-v^{(2)}), $$ $$
\  R^2_{12}(u^{(1)},u^{(2)}| v^{(2)}) = \mathcal{S}_{12}(u^{(2)}-u^{(1)})
\mathcal{S}_{11}(u^{(2)}-v^{(2)})   \mathcal{S}_{12}(u^{(1)}-v^{(2)}).
$$
This leads to  the expressions
\begin{align*}
&R^1_{12}(u^{(1)}|v^{(1)}, v^{(2)}) = \frac{\Gamma(x_{21} \dd_2 +
u^{(1)}-v^{(2)}+1)}{\Gamma(x_{21} \dd_2+v^{(1)}-v^{(2)}+1)}, \\
& R^2_{12}(u^{(1)}, u^{(2)}| v^{(2)}) = \frac{\Gamma(x_{12} \dd_1 +
u^{(1)}-v^{(2)}+1)}{\Gamma(x_{12} \dd_1+u^{(1)}-u^{(2)}+1)},
\end{align*} 
where $ x_{12} = x_1-x_2$ and $x_{21} = x_2- x_1$. 

Whereas the action of the  operator $\mathcal{S}_{12}(u) $ on functions of
$ x_1,x_2 $
 is multiplicative, the action of the latter operators and of $ \mathcal{S}_{11} $,
$\mathcal{S}_{22} $ can be defined with the help of
the Beta integral representation and $\hat N = x \dd$. We have
$$W(2\ell+1) = \frac{x^{-2\ell-1}}{\Gamma(-2\ell-1)} B(\hat N+1, -2\ell-1)
= \mathrm{const} \frac{1}{x^{2\ell+1}}\int \mathrm{d}s\, s^{\hat N}  (1-s)^{-2\ell -2}. $$
The action on functions can be defined (up to a constant factor) as
\be \label{Wf}
 W(2\ell+1) f(x) = 
\frac{1}{x^{2\ell+1}}\int \mathrm{d}s\,   (1-s)^{-2\ell -2} f(sx) =
 \int \mathrm{d}\tilde x \,   (x-\tilde x)^{-2\ell -2} f(\tilde x). 
\ee
This is clear by $s^{x\dd} x = x s^{x\dd +1} = (sx) s^{x\dd} $.

We shall consider the transformation of monodromy matrices like
$ T(\mathbf{v}) R = R T(\mathbf{v}\p) $ accompanied by
permutations $\tilde \sigma$  of the set of $N$ parameter pairs 
$\mathbf{v} = v_1^{(1)}, v_1^{(2)}; \dots; v_N^{(1)}, v_N^{(2)}. $ 
The resulting parameter set is now conveniently presented 
as one string of $N$ pairs
$$  v_{\sigma(1^{(1)})}, v_{\sigma(1^{(2)})}; \dots; v_{\sigma(N^{(1)})},
v_{\sigma (N^{(2)})}. $$

%%%%%%%%%%%%%%%%%%%%%%%%%%%%%%%%%%%%%%%%%%%%%%%%%%%%%%%%%%%%

Restricting the $L$ operator in the helicity form to a subspace of a
particular helicity eigenvalue $2h$ of 
$-\dd^{\lambda} \lambda + \bar \lambda \bar \dd^{\lambda} $ 
related to the coordinate dilatation weight  $2\ell$ as $2h = 2\ell +1$, 
we may introduce another
form of the $L$ operator involving only one canonical pair.
We change from $\lambda, \bar \lambda$ to $k, l$
 and calculate  the corresponding conjugated
operators from
$$ \bar \lambda^2= k l, \quad  \lambda^2= \frac{k}{l}, \quad  \lambda \dd^{\lambda} =
k \dd_k - l \dd_l, \quad
 \bar \lambda \bar \dd_{ \lambda} = k \dd_k + l \dd_l. $$
We  restrict the  
functions of $\lambda, \bar \lambda$  to   the form 
$\psi(\lambda, \bar \lambda) = l^h \phi(k) $. 
$$ L^{\lambda}(u) \cdot l^h \phi(k)  = l^h L^k(u+h,h) \cdot  \phi(k) $$ 
and  obtain $L^k(u,h) = I u + L^k(0,h)$, 
$$ L^k(0,h) = \begin{pmatrix}
 - k\dd_k  & -k \\
\frac{1}{k} (k\dd_k+h) (k\dd_k -h) & \dd_k k 
\end{pmatrix}.
$$
The result is to be compared with the ratio coordinate reduction
$L(v^{(1)}, v^{(2)})$ after the canonical transformation
$x, \dd \to \dd_k, -k$. 
If the ratio coordinate $x$ describes  position 
then $k$ plays the role of  momentum.  

We add the remark that the factor carrying the weight can be chosen
differently, e.g., $\psi(\lambda, \bar \lambda) = l^h k^a \phi\p(k) $. 
 Then the equivalent representation of the algebra on $\phi\p(k)$ is generated by
$L^{k \prime}(0,h) = k^{-a} L^k(0,h) k^a $.

\subsection{Constructions of symmetric correlators}

Any transformation of the   monodromy matrix $T \to T\p$ induces a map from a
solution $\Phi$ of the symmetry condition  (\ref{YSC}) with $T$ 
to a solution $\Phi\p$ of the
symmetry condition with $T\p$. Examples of such transformations have been
discussed in \cite{CK}. They are related to features observed in the
construction of scattering amplitudes in gauge field theories \cite{AH}.

If we  separate the set of points into two subsets $ 1, \dots, N \to 1, \dots, M;
M+1,\dots, N $ and correspondingly factorize the monodromy matrix
$T_{1,\dots.,N} = T_{1,\dots,M} T_{M+1, \dots,M} $
the symmetric correlator $\Phi^{(1)}(\mathbf{x}_1,\dots, \mathbf{x}_M)$
of the condition (\ref{YSC}) with $ T_{1,\dots,M}$ and the symmetric correlator
$\Phi^{(2)}(\mathbf{x}_{M+1}, \dots, \mathbf{x}_N)$  
of the condition (\ref{YSC}) with $ T_{M+1,\dots,N}$  results in a solution of the
condition (\ref{YSC}) with the full monodromy matrix $T_{1,\dots,N}$ simply by
product, $ \Phi = 
\Phi^{(1)}(\mathbf{x}_1,\dots., \mathbf{x}_M) \ 
\Phi^{(2)}(\mathbf{x}_{M+1},\dots, \mathbf{x}_N)$.

We find immediately elementary correlators solving the symmetry condition
in the case $N=1$ where the monodromy consists of one factor $L$ only.
$$ L^+(u) \cdot 1 = I\, (u+1), \ \ \  L^- (u)\cdot 1  =I\,  u,  $$
\be \label{L+-1}
 L^+(u) \cdot \delta^{(2)}(\mathbf{x})  = I \,  u \delta^{(2)}(\mathbf{x}), \ \ \ L^- (u)\cdot
\delta^{(2)}(\mathbf{x})  = I \, (u+1) \delta^{(2)}(\mathbf{x}).\ee
In the first case the weight is $2\ell = 0$, in the second  $2\ell = -2$.
\be \label{Llambdel}
 L^{\lambda}(u) \cdot \delta(\lambda)= I \, (u+1) \delta(\lambda), \ \ \  
L^{\lambda}(u)\cdot \delta(\bar \lambda)  = I \, u \delta(\bar \lambda). \ee
Here the weights are again $0$ and $-2$, correspondingly. The weights are
calculated  as eigenvalues of
$ (\mathbf{x} \cdot \mathbf{p}) = \bar \lambda \bar \dd^{\lambda} - 
\dd^{\lambda} \lambda  $. In ratio coordinates we have
\be \label{Lv1v21}
  L(v, v-1) \cdot 1  = I \,  v  \ee
In the last  case the weight is zero  and therefore the two arguments are
related as $v^{(1)} - v^{(2)} = 1$.

Starting from a solution of the Yangian symmetry condition (\ref{YSC})
other solutions can be generated by action on $\Phi$ with 
YB operators obeying particular $RLL$ relations with 
$L$ operators entering the monodromy. 
This can be implemented both in the formulations with homogeneous
coordinates  or with coordinate ratios. It is an important tool
of constructing YSC and it will be illustrated in the following section. 

We act with a sequence of $R$ operators on a trivial correlator composed as
a product of elementary 1-point correlators given above. The $R$ operators,
in particular the points on which they act non-trivially and their
arguments, have to be chosen such that the $RLL$ relations allow to permute
them with the product of $L$ operators of the monodromy matrix.
In this way we obtain that the particular sequence of $R$ operations applied
to the initial trivial correlator  produces another Yangian symmetric
correlator corresponding to the monodromy resulting by the permutations 
from the initial monodromy. 

In the form (\ref{Sa}) the generators correspond to infinitesimal Moebius
transformations, where the finite transformation can be written as
$x \to x\p = \frac{a x +b}{cx +d}$, $ad-bc= 1 $,  in the particular case of
vanishing weight, $2\ell=0$. The conformal transformations acting on functions  as
$$ F(x_1, \dots, x_N) \to \prod (cx_i + d)^{2\ell_i} \ F(x_1\p, \dots,
x_N\p) $$
are generated by $\sum S_i^a$. Conformal symmetric correlators are invariant
under this transformation and thus obey the condition 
\be \label{conformal}
\sum_1^N S_i^a \cdot  \Phi(x_1, \dots, x_N) = 0. \ee
The general form of $N$-point functions obeying (\ref{conformal})
is well known. Beyond $N=3$ the condition admits an arbitrary dependence on
anharmonic ratios like $\frac{x_{12} x_{34}}{x_{13}x_{24}}$. 
The Yangian symmetry condition (\ref{YSC}) implies (\ref{conformal}).
Moreover, it fixes the freedom left by the global conformal symmetry.

\section{Examples of Yangian symmetric correlators} \label{sec:3}
\setcounter{equation}{0}

If working in the homogeneous coordinate form each $R^{++}$ operator action comes
with a contour integral (\eqref{R12}, Appendix \ref{app:A}). The resulting multiple
integral can be rewritten in the standard link integral form with the
integration variables being coordinates of the related Grassmannian. 
This step is not specific to $s\ell_2$ and the resulting expressions coincide with
corresponding ones for amplitude calculations ($s\ell_4$) \cite{AH}.
The specifics of the $s\ell_2$  case appears in the next steps to obtain the
explicit forms in homogeneous, helicity or ratio coordinates. We have also
the alternative to work with the $R$ operators factors (\ref{S12}) in the ratio
coordinate form directly.

\subsection{Two-point correlators }
 
We start with two-point correlators obtained from a trivial one by just one
$R$ operation. 
This can be formulated in homogeneous coordinates with
$R^{++}$ (\ref{R12}), in helicity coordinates with $R^{\lambda}$,
in homogeneous coordinates with $R^{+-}$  or directly in ratio
coordinates with $\mathcal{S}_{12}$ (\ref{S12}).

\vspace{.5cm}

{\it Homogeneous coordinates}

\noindent
The two-point YSC with the monodromy 
$T_2= L_1^+(u^+_2,u_1) L_2^+(u_1^+,u_2)$, where $u_1^+=u_1-2$ and $u_2^+=u_2$, is
$$\Phi^{-+}(u_1,u_2)  = R^{++}_{21}(u_1^+-u_2^+) \cdot \delta^{(2)}(\mathbf{x}_1)
= 
\int \frac{\mathrm{d}c}{c^{1+u_1^+- u^+_2}} \delta^{(2)} (\mathbf{x}_1 -
c \mathbf{x}_2 ).$$
The monodromy results from the initial one $T^{(0)}_2 = L_1^+(u_1^+,u_1)
L_2^+(u_2^+,u_2)$ by the permutation of the first parameters $u_1^+ = u_1-2,
u_2^+ = u_2$. This is written conveniently by the permutation pattern
$$ \begin{pmatrix}
1  & 2 \\
2  & 1
\end{pmatrix}.
$$
We do the integration with one of the two delta distributions and 
transform the other one.
$$ \Phi^{-+}(u_1,u_2) = \left (\frac{x_{2,2}}{x_{1,2}} \right )^{1+ u_1^+ -
u_2^+} \delta(\langle 12 \rangle)  
= \left(\frac{x_{2,2}}{x_{1,2}}\right)^{1+2 \ell_2} \frac{1 }{x_{1,2} x_{2,2}} \delta(x_1 -
x_2). $$
We use the notation
$$ \langle 12 \rangle = x_{1,1} x_{2,2} - x_{1,2} x_{2,1}, \ \ \ x_1 =
\frac{x_{1,1}}{x_{1,2}}, \ \ \  x_2 =\frac{x_{2,1}}{x_{2,2}}. $$ 
We express the exponents in terms of the weights
$$ 2\ell_1 = u_2^+ - u_1, \ \  2\ell_2 = u_1^+ - u_2, \ \  2\ell_1 + 2\ell_2 +2 = 0 $$
and obtain
$$\Phi^{-+}(u_1,u_2)  = x_{1,2}^{2\ell_1}  x_{2,2}^{2\ell_2} \delta(x_1 -      
x_2). $$
The weights appear as degrees of homogeneity as expected. 

\vspace{.5cm}

{\it Helicity coordinates}

\noindent
With the same monodromy now written  in $L^{\lambda}$ we have the
correlator in the helicity form
$$\Phi^{-+}(u_1,u_2)  = R^{\lambda} (u_1^+ -u^+_2) \cdot \delta(\bar \lambda_1)
\delta(  \lambda_2) = 
\int \frac{\mathrm{d}c}{c^{1+u_1^+- u^+_2}} 
\delta (\bar \lambda_1 - c \bar \lambda_2 ) \delta ( \lambda_2 + c
 \lambda_1 ) = $$ $$
= \bar \lambda_1^{1+ 2\ell_1} \bar \lambda_2^{1+ 2\ell_2} \delta(\lambda_1 \bar
\lambda_1 + \lambda_2 \bar \lambda_2). $$ 
The result appears as a prefactor with exponents $2h= 2\ell+1$ 
and a scale independent factor. 

\vspace{.5cm}

{\it Ratio coordinates }

\noindent
Another two-point correlator with the monodromy $T_2=
L^+_1(u^+_1,u_2)L^-_2(u_1, u_2^-)$, where $u_1^+=u_1$ and $u_2^-=u_2-2$, is
$$ \tilde \Phi = R_{12}^{+-}(u_1-u_2) \cdot 1 = (\mathbf{x}_1
\mathbf{x}_2)^{u_1-u_2} =
 (x_{1,2} \ x_{2,1})^{u_1-u_2} \ (x_1-x_2)^{u_1-u_2}.
$$
The second factor in the last form is the two-point correlator obtained 
directly in the framework of the ratio-coordinate form by the action of
$\mathcal{S}_{12}$. 
In the notation introduced for the ratio coordinate form the initial 
monodromy reads here $T_2^{(0)}= L_1(v_1, v_1-1) L_2(v_2,v_2-1)$ and the
action by  $\mathcal{S}_{12}$ results in the permutation of the second
parameter of the first factor with the first of the second,
$T_2 = L_1(v_1, v_2) L_2(v_1-1,v_2-1)$. This is conveniently written by the
parameter string displaying just the sequence of arguments.  
Indeed, the correlator with the monodromy $T_2$ (characterized by the 
corresponding parameter string) is
$$ (x_1-x_2)^{v_1-v_2-1}, \ \ \ T_2: v_1,v_2; v_1-1,v_2-1. $$

\subsection{ Three-point correlators } 

We present examples of three-point correlator constructions with
representations in homogeneous, helicity and ratio coordinates.
The direct construction in ratio coordinates by three steps of action with
the elementary permutation operators $\mathcal{S}_{ij}$ is described.

\vspace{.5cm}

{\it Homogeneous coordinates }

\noindent
The three-point YSC with the monodromy characterized by the
parameter permutation pattern
$$
\begin{pmatrix}
1&2&3 \\
2&3&1
\end{pmatrix}
$$ obtained by the action with $R_{21}^{++} R_{31}^{++}$ is 
$$  \Phi^{-++} = R_{21}^{++}(u^+_3-u^+_2) R_{31}^{++}(u_1^+-u_3^+) \cdot
\delta^{(2)}(\mathbf{x}_1 ) = 
\int \mathrm{d}^2c \, \varphi_3 \, \delta^{(2)}(\mathbf{x}_1 -c_{12}
\mathbf{x}_2 - c_{13} \mathbf{x}_3),  $$ \vspace{-10pt}
$$ \varphi_3^{-1} = c_{12}^{1+ u_3 - u_2} c_{13}^{1+ u_1^+ - u_3}.  $$
There are two delta distributions with the arguments
$$ x_{1,1} - c_{12} x_{2,1}- c_{13} x_{3,1} = x_{1,1}\p , \ \ \
x_{1,2} - c_{12} x_{2,2}- c_{13} x_{3,2} = x_{1,2}\p.
$$
The equations $x_{1,1}\p = 0$, $x_{1,2}\p = 0 $ are solved by
$$ c_{12}^{(0)} = \frac{ \langle 13 \rangle}{\langle 23 \rangle }, \ \ \ c_{13}^{(0)} = \frac{
\langle 12 \rangle}{\langle 32 \rangle},
$$
and the Jacobi determinant is
$$ \frac{\dd (x_{1,1}\p, x_{1,2}\p )}{\dd (c_{12}, c_{13})} = \langle 23 \rangle. $$
From the permutation pattern we express the exponents in terms of the weights, 
$$ 2\ell_1 = u_2 - u_1, \quad 2\ell_2 = u_3 - u_2, \quad 2\ell_1 + 2\ell_2 + 2\ell_3
+ 2 = 0, $$
and obtain the result
$$ \Phi^{-++} = 
\frac{\langle 23 \rangle^{1+ 2\ell_2 + 2\ell_3} }{\langle 13 \rangle^{1+2\ell_2} \langle 12\rangle^{1+ 2\ell_3} }.
$$
 The symplectic products $\langle ij \rangle$
can be substituted by the differences in the ratio coordinates
$x_{ij}= x_i - x_j, x_i = \frac{x_{i,1}}{x_{i,2}} $, 
$$ \Phi^{-++} = x_{1,2}^{2 \ell_1} x_{2,2}^{2\ell_2} x_{3,2}^{2\ell_3} \ 
\frac{x_{2 3}^{1+ 2\ell_2 + 2\ell_3} }{x_{1 3}^{1+2\ell_2} x_{1 2}^{1+ 2\ell_3
} }.
$$

\vspace{.5cm}

{\it Helicity coordinates}

\noindent
The  three-point correlator with the same monodromy written in the helicity
form in terms of $L^{\lambda}$
is  ($k_i= \lambda_i \bar \lambda_i$)
 \begin{multline*} \Phi^{-++} = \int \mathrm{d}^2c \, \phi^{-++} \, 
\delta(\bar \lambda_1 -c_{12}\bar \lambda_2 - c_{13} \bar \lambda_3)
\delta( \lambda_2 +c_{12} \lambda_1) 
\delta( \lambda_3 + c_{13}  \lambda_1)
\\
=\left (\frac{\bar \lambda_1}{ \lambda_1} \right)^{h_1}
\left (\frac{\bar \lambda_2}{ \lambda_2} \right)^{h_2}
\left (\frac{\bar \lambda_3}{\lambda_3} \right)^{h_3}
\ k_1^{-h_1} k_2^{-h_2}k_3^{-h_3}
\ \delta(k_1+k_2+k_3).
\end{multline*}

\vspace{.5cm}

{\it Ratio coordinates} 

\noindent
 We add examples obtained in the framework of the ratio-coordinate form.
We start from the monodromy matrix in ratio coordinates and the trivial
constant correlator
$$ L_1(v_1,v_1-1) L_2(v_2,v_2-1)L_3(v_3,v_3-1) \cdot 1 = I\, v_1 v_2 v_3 . $$
We abbreviate the monodromy matrix by the string of parameter pairs:
$T_3:v_1,v_1-1;v_2,v_2-1;v_3,v_3-1$. 
As the first step we find the symmetric correlator with the monodromy
$$\mathcal{S}_{13} (v_1^{(2)}-v_3^{(1)} ) = (x_1-x_3)^{v_1-v_3-1}, \ \ \
T_3: v_1,v_3;v_2,v_2-1;v_1-1,v_3-1. $$
 Because of $L_2$ acting on a constant results
in the unit matrix the $RLL$ relation with $\mathcal{S}_{13}$
for the product of $L_1 L_3$ applies here.
Next we apply the $RLL$ relation with $\mathcal{S}_{12}$ and find 
the correlator with the monodromy
$$ (x_1-x_3)^{v_1-v_3-1} \ (x_1-x_2)^{v_3-v_2}, \ \ \
T_3:v_1,v_2;v_3,v_2-1;v_1-1,v_3-1.
$$
 We apply now the $RLL$ relation
with $\mathcal{S}_{23}$ to establish the 
symmetric correlator with the monodromy of the parameter string
\be \label{YSCS12}
 (x_1-x_3)^{v_1-v_3-1} \ (x_1-x_2)^{v_3-v_2} \ (x_2-x_3)^{v_2-v_1}, \ \ \ 
T_3: v_1,v_2;v_3,v_1-1;v_2-1,v_3-1. \ee

\subsection{4-point correlators}

We consider again examples of construction in homogeneous, helicity and
ratio coordinate form. The results will be used in the following to
define symmetric integral operators and to establish relations to the
QCD parton evolution. 

\vspace{.5cm}

{\it Homogeneous coordinates}

\noindent
Let us consider the example
\begin{align*}
\Phi^{--++} = R^{++}_{43}(u_2^+ - u_1^+) R^{++}_{32}(u_2^+ - u_4^+) 
R^{++}_{43} (u_1^+ - u_4^+)
R^{++}_{31}(u_1^+ -u_3^+) \delta^{(2)} (\mathbf{x}_1) \delta^{(2)} (\mathbf{x}_2)  \\ \rule{0pt}{20pt}
=\int \frac{dc_{34}^{(2)} dc_{23} dc_{34}^{(1)}dc_{13}}
{c_{34}^{(2) 1 + u_2^+ - u_1^+} c_{23}^{ 1 + u_2^+ - u_4^+}
c_{34}^{(1) 1 + u_1^+ - u_4^+} c_{13}^{ 1 + u_1^+ - u_3^+} }
\delta^{(2)}(\mathbf{x}_1 - c_{13} \mathbf{x}_3 + c_{13}(c_{34}^{(1)}+
c_{34}^{(2)})\mathbf{x}_4) \\
\qquad \qquad \qquad\qquad \qquad \qquad\qquad \qquad \qquad \qquad \quad \times \delta^{(2)} (\mathbf{x}_2  - c_{23} \mathbf{x}_3 + c_{23}
c_{34}^{(2)}\mathbf{x}_4 ).
\end{align*} 
The permutation pattern is
$$
\begin{pmatrix}
1 & 2 & 3 & 4 \\
3 & 4 & 1 & 2   
\end{pmatrix}.   
$$
We transform to the normal link integral form, where the integration variables
coincide with the coefficients in the arguments of the $\delta$-functions. This
 can be formulated in terms of the link matrices
\be \label{cmatrix}
\begin{pmatrix}
1 & 0 & -c_{13} & c_{13}(c_{34}^{(1)}+ c_{34}^{(2)}) \\
0 & 1 & -c_{23} &    c_{23}   c_{34}^{(2)}
\end{pmatrix}  
\ = \ 
\begin{pmatrix}
1 & 0 & - c_{13} & - c_{14} \\
0 & 1 & - c_{23} & -  c_{24}  
\end{pmatrix}.  
\ee
We obtain
$$ c_{34}^{(2)} = - \frac{ c_{24}}{c_{23}}, \ \ 
c_{34}^{(1)} = \frac{c_{13}  c_{24} -  c_{14} c_{23} }{c_{13}
c_{23} }, $$
and the normal link integral form is 
\be \label{link1}
 \Phi^{--++} = \int \mathrm{d}^4c \, \varphi_4 \,
\delta^{(2)}(\mathbf{x}_1 -c_{13}\mathbf{x}_3 - c_{14} \mathbf{x}_4) 
\delta^{(2)}(\mathbf{x}_2 -c_{23}\mathbf{x}_3 - c_{24} \mathbf{x}_4),  
\ee \vspace{-10pt}
$$ \varphi^{-1}_4 = c_{13}^{1+ u_4 - u_3} c_{24}^{1+ u_2^+ - u_1^+}  
(c_{13} c_{24} - c_{14} c_{23})^{1+ u_1^+ - u_4}. 
$$
We have four delta distributions.
The vanishing of the arguments of the first two result in
$$ c_{13}^{(0)} = \frac{\langle 14 \rangle }{\langle 34 \rangle}, \ \ \  c_{14}^{(0)} =
\frac{\langle 13 \rangle}{ \langle 43 \rangle}, $$
and of the last two
$$ c_{23}^{(0)} = \frac{\langle 24 \rangle}{\langle 34 \rangle}, \ \ \  c_{24}^{(0)} =
\frac{\langle 23 \rangle}{\langle 43 \rangle}.$$
Further,
$$c^{(0)}_{13} c^{(0)}_{24} - c^{(0)}_{14} c^{(0)}_{23} = \frac{\langle 12 \rangle}{\langle 34 \rangle}. $$
The Jacobi determinant is
$$ \frac{\dd (x_{1,1}\p, x_{1,2}\p, x_{2,1}\p, x_{2,2}\p )}
{\dd (c_{13}, c_{14},c_{23}, c_{24}) } = \langle 34\rangle^2. $$
The result is 
$$ \Phi^{--++} = \langle 34\rangle^{-2} \varphi^{--++}(c^{(0)}) =  
\frac{\langle 12 \rangle^{1+ u_4 - u_1} }{\langle 14 \rangle^{1+u_4-u_3} \langle 23 \rangle^{1+ u_2 - u_1} 
\langle 34\rangle^{1+u_3 -u_2} }.
$$
The exponents can be expressed in terms of the weights 
\be \label{l1l2}
 2\ell_1 = u_3 - u_1,\ \  2\ell_2 = u_4 - u_2, \ \ 
2\ell_1 + 2\ell_3 + 2 = 0,\ \ 2\ell_2 + 2\ell_4 + 2 = 0, \ee
and further the weights can be replaced by the helicities
 $2h =  2 \ell +1$. Because only two of the weights are independent, there appears
one   difference in the spectral parameters additionally
$$ \Phi^{--++} =  
\frac{\langle 12 \rangle^{1+ 2 \ell_1 +2\ell_2} }{\langle 14 \rangle^{1+2\ell_2} \langle 23 \rangle^{1+ 2\ell_1} 
\langle 34\rangle }
\left ( \frac{\langle 12 \rangle \langle 34\rangle }{ \langle 14 \rangle \langle 23 \rangle } \right )^{u_2-u_3}. $$
The appearance of additional parameters was noticed in the amplitude context
and proposed as a tool of regularization \cite{FLMPS13}. 

The result depends homogeneously on the components of the points $i$
with  the weights $2\ell_i$. The symplectic products $\langle ij \rangle$ can be
replaced by the differences in the relative coordinates 
$x_{ij} = x_i - x_j $, $x_i = \frac{x_{i,1} }{x_{i,2}}$
\be \label{4corr1}
 \Phi^{--++} =  
x_{1,2}^{2\ell_1} x_{2,2}^{2\ell_2} x_{3,2}^{2\ell_3} x_{4,2}^{2\ell_4} 
\frac{x_{1 2}^{1+ 2 \ell_1 +2\ell_2 }}{x_{1 4}^{1+2\ell_2} x_{2 3}^{1+ 2\ell_1} 
x_{3 4} }
\left ( \frac{x_{1 2}  x_{3 4} }{ x_{1 4} x_{2 3} } \right )^{u_2-u_3}. \ee

\vspace{0.5cm}

{\it Stepwise construction for kernels} \\
\noindent
We consider a way to an equivalent correlator 
where the intermediate steps have particular
interpretations as kernels (see section \ref{sec:5.2}). 
$$  \Phi^{--++} = 
R^{++}_{12}(u_1-u_2) R^{++}_{21}(u_4-u_3) R^{++}_{32}(u_2^+ - u_3) 
R^{++}_{41} (u_1^+ - u_4)
\delta^{(2)}(\mathbf{x}_1) \delta^{(2)}(\mathbf{x}_2). $$
The first two $R$ operations result in
\begin{multline} \label{Delta}
 R^{++}_{32}(u_2^+ - u_3) R^{++}_{41} (u_1^+ - u_4)
\delta^{(2)}(\mathbf{x}_1) \delta^{(2)}(\mathbf{x}_2) \\
=\int \frac{\mathrm{d}c_{23} \mathrm{d}c_{14}
}{
c_{23}^{1+ u_2^+ - u_3}  c_{14}^{1+u_1^+ - u_4} }
\delta^{(2)}(\mathbf{x}_1- c_{14} \mathbf{x}_4) \delta^{(2)}(\mathbf{x}_2 - c_{23}
\mathbf{x}_3) = \Delta^{--++} (1,2,3,4). \end{multline}
The monodromy is characterized by the permutation pattern
$$ \begin{pmatrix}
1&2&3&4 \\
4&3&2&1   
\end{pmatrix}. $$
In the next step we have
\begin{multline*}
R^{++}_{21}(u_4-u_3) \Delta^{--++}  =
\int \frac{\mathrm{d}\bar c_{13} \mathrm{d}c_{23} \mathrm{d}c_{14} }{\bar c_{13}^{1+ u_4-u_3}
c_{23}^{1+ u_2^+ -  u_4}  c_{14}^{1+u_1^+ - u_4} }
\delta^{(2)}(\mathbf{x}_1-  \bar c_{13} \mathbf{x}_3-c_{14} \mathbf{x}_4) \\
\times \delta^{(2)}(\mathbf{x}_2 - c_{23}\mathbf{x}_3),
\end{multline*} 
where $\bar c_{13} $ results from a transformation of the integration variables
coming with the $R$ operators.
This result is a YSC with the monodromy characterized by
$$ \begin{pmatrix}
1&2&3&4 \\
3&4&2&1   
\end{pmatrix}. $$
After the fourth $R$ action the integration variables form the matrix
$$ \begin{pmatrix}
1 & 0 & c_{12} c_{23} & c_{14} \\
0 & 1 & c_{23} (1+c_{21} c_{12}) &  c_{21} c_{14}
\end{pmatrix}
= 
 \begin{pmatrix}
1 & 0 & \bar c_{13}  & \bar c_{14} \\
0 & 1 & \bar c_{23}  &  \bar c_{24}  
\end{pmatrix},
$$
from which the transformation to the normal link form can be read off. 
We notice that the Jacobian is $c_{14} c_{23}$ and that
the resulting power of $\bar c_{14}$ vanishes. We  obtain
for the connected 4-point correlator
\begin{multline*}
\Phi^{--++} = \int \frac{\mathrm{d}\bar c_{13} \mathrm{d}\bar c_{14} \mathrm{d}\bar c_{23} \mathrm{d}\bar
c_{24} }{
\bar c_{13}^{1+ u_4-u_3} \bar c_{24}^{1+ u_1 - u_2} (\bar c_{23} \bar
c_{14} - \bar c_{13}\bar c_{24})^{1+ u_2^+ - u_4} } 
\delta^{(2)} (\mathbf{x}_1 - \bar c_{13} \mathbf{x}_3 - \bar c_{14}\mathbf{x}_4) \\
\times \delta^{(2)} (\mathbf{x}_2  - \bar c_{23} \mathbf{x}_3 - \bar c_{24}\mathbf{x}_4
).
\end{multline*} 
The resulting permutation pattern allows to calculate the weights,
$$ 2\ell_1 = u_3-u_2,\ \ 2\ell_2 = u_4 - u_1,\ \  2\ell_1 + 2\ell_3 +2=0, \ \ 
2\ell_2 + 2\ell_4 +2 = 0. $$
The exponents in the result can be expressed in terms of the weights and
a difference of two spectral parameters,
$$ u_1 - u_2 = 2\ell_1 - 2\ell_2 - (u_3-u_4), \ \ u_2^+ - u_4 = -2\ell_1 - 2 +
(u_3 - u_4), $$
\be \label{kernel}
 \Phi^{--++} =  
x_{1,2}^{2\ell_1} x_{2,2}^{2\ell_2} x_{3,2}^{2\ell_3} x_{4,2}^{2\ell_4}
\, \cdot \, 
\frac{x_{1 2}^{1+ 2 \ell_1  }}{x_{14} x_{23}^{1+ 2\ell_1-2\ell_2} 
x_{34}^{1+2\ell_2} }
\left (\frac{x_{12}  x_{34} }{ x_{14} x_{23} } \right )^{u_4-u_3}. 
\ee
The first factors carrying the weights $2\ell_i$ are written in the second
component of the homogeneous coordinates, but the last factors are
written in the differences of the ratio coordinates, i.e. $
 x_{12} = \frac{x_{1,1}}{x_{1,2}} - \frac{x_{2,1}}{x_{2,2}}$.
Coincidence with the result of the first way (\ref{4corr1})
is obtained by the substitution of the parameters as
\be \label{usub1}
 u_4 - u_3 \to u_2 - u_3 + 2\ell_2. \ee

\vspace{0.5cm}

{\it Helicity representation}\\
\noindent
The transformation  to the helicity representation is easily done in the
integral over the link variables $c$ (\ref{link1})
by modifying the delta factors in the integrand
\begin{multline*}
\Phi^{--++} = \int \mathrm{d}^4c \, \varphi_4 \, 
\delta(\bar\lambda_1-c_{13}\bar \lambda_3-c_{14}\bar\lambda_4) 
\delta(\bar \lambda_2-c_{23}\bar \lambda_3-c_{24} \bar\lambda_4) 
\delta(\lambda_3 +c_{13}\lambda_1 + c_{23}\lambda_2) \\
\times \delta(\lambda_4 +c_{14}\lambda_1 + c_{24}\lambda_2). 
\end{multline*} 
We have again four delta distributions with the arguments 
$$\bar \lambda_1 -c_{13}\bar \lambda_3 - c_{14}\bar \lambda_4 =\bar \lambda_1\p , \quad 
\bar \lambda_2 -c_{23}\bar \lambda_3 - c_{24}\bar \lambda_4 = \bar \lambda_2\p, $$ $$
 \lambda_3 +c_{13} \lambda_1 + c_{23}  \lambda_2 = \lambda_3\p , \quad
 \lambda_4 +c_{14} \lambda_1 + c_{24}  \lambda_2 = \lambda_4\p, 
$$
but they are not removing the
four link integrations. We see the dependence between the related 
linear equations by multiplying the first by $ \lambda_1$, the second by
$ \lambda_2$, the third by $\bar \lambda_3$, the fourth by $\bar \lambda_4$, and
adding them with the result
$$ \lambda_1 \bar \lambda_1 + 
\lambda_2 \bar \lambda_2 + \lambda_3 \bar \lambda_3
+\lambda_4 \bar \lambda_4 = 0. $$
We use the first three $\delta$-functions to do the integrations over $c_{13}, c_{14},
c_{23}$. The Jacobi factor is
$$ \frac{\dd(\bar\lambda_1\p, \bar\lambda_2\p, \lambda_3\p)}{\dd(c_{13}, c_{14},
c_{23})} = \bar\lambda_3 \bar\lambda_4  \lambda_1. $$ 
Their values  are fixed at
$$ c_{13}^{(0)} = \frac{\lambda_1 \bar \lambda_1 +\lambda_4 \bar \lambda_4 +
c_{24} \bar \lambda_4  \lambda_2 }{ \bar\lambda_3  \lambda_1}, \ \ \ 
c_{14}^{(0)} = \frac{- \lambda_4 - c_{24}  \lambda_2}{
\lambda_1}, $$  $$
c_{23}^{(0)} = \frac{\bar \lambda_2 - c_{24} \bar \lambda_4}{\bar
\lambda_3}, \ \ \ 
c_{13}^{(0)}c_{24}- c_{14}^{(0)}c_{23}^{(0)}
 = \frac{ \bar \lambda_2  \lambda_4 + c_{24} (\lambda_1 \bar \lambda_1
+\lambda_2 \bar \lambda_2)}{\bar \lambda_3  \lambda_1}. $$
We transform the fourth delta distribution by taking it together with the
third one. The transformation should multiply the matrix of coefficients
$$
\begin{pmatrix}
c_{13} & c_{23} & 1 & 0 \\
c_{14} & c_{24} & 0 & 1
\end{pmatrix}
$$
by the matrix 
$$ A= 
\begin{pmatrix}
1 & 0  \\
\bar\lambda_3 & \bar\lambda_4
\end{pmatrix}
$$
to bring the second row into the wanted form. Then we get
$$ \delta( \lambda_3\p) \delta(\lambda_4\p) = \det (A)
\, \delta( \lambda_3\p) \, \delta (\lambda_1 \bar \lambda_1 + 
\lambda_2 \bar \lambda_2 + \lambda_3 \bar \lambda_3
+\lambda_4 \bar \lambda_4). $$
The resulting form of the correlator is
 $$ \Phi^{--++} = \int \mathrm{d}c_{24} \, \varphi_4(c^{(0)}(c_{24}) ) \,
\frac{1}{ \lambda_1 \bar \lambda_3} \,
\delta (\lambda_1 \bar \lambda_1 + 
\lambda_2 \bar \lambda_2 + \lambda_3 \bar \lambda_3
+\lambda_4 \bar \lambda_4), $$
where $\varphi_4 $ is given in (\ref{link1}). 
We change to
$ c = c_{24}  \lambda_2 \bar \lambda_4 $
and $ k_i = \lambda_i \bar \lambda_i$ in order to separate the
one-dimensional momentum dependence from the homogeneity factor.
\begin{multline*} \Phi^{--++} = \lambda_1^{u_1-u_3-1} \lambda_2^{u_2-u_4 -1} 
\bar \lambda_3^{u_1-u_3-1} \bar \lambda_4^{u_2-u_4 -1}
\ \delta(k_1+k_2+k_3+ k_4) \\
\times \int \mathrm{d}c \, c^{u_1-u_2-1} \ (k_1+k_4+c)^{u_3-u_4-1} \
(k_2k_4+c(k_1+k_2))^{u_4-u_1+1}.
\end{multline*}
We substitute the weights owing to the permutation pattern as in (\ref{l1l2})
and use the helicity notation $2h = 2\ell+1$. Recall that we have here only two
independent weights and we choose $\eps = u_2-u_3$ as the extra
parameter.
The integration over $c$ is simplest in the case of vanishing
momentum transfer in the channel $12 \to 34$ to which we shall refer later.
\begin{multline} \label{4corrk}
 \Phi^{--++}|_{k_1+k_2 \to 0}  = \lambda_1^{u_1-u_3-1} \lambda_2^{u_2-u_4 -1} \bar
\lambda_3^{u_1-u_3-1} \bar \lambda_4^{u_2-u_4 -1}
\, \delta(k_1+k_2+k_3+ k_4) \\
\times \mathrm{const}\, (k_1 k_3)^{2h_1+2h_2+ \eps-1 } (k_1-k_3)^{-2h_1-2h_2 -2\eps+1}.
\end{multline}

\vspace{0.5cm}

{\it Ratio-coordinate construction} \\
\noindent
We add an example of construction in the ratio-coordinate form.
\begin{multline*}
\bar \Phi = \mathcal{S}_{12}(v_4^{(1)},v_2^{(1)})
\mathcal{S}_{34}(v_3^{(2)},v_1^{(2)})\mathcal{S}_{23}(v_2^{(2)},v_3^{(1)})
\mathcal{S}_{14}(v_1^{(2)},v_4^{(1)}) \cdot 1 \\
=x_{23}^{v_2-v_3-1} x_{14}^{v_1-v_4-1} x_{12}^{v_4-v_2} x_{34}^{v_3-v_1}. 
\end{multline*}
The resulting monodromy is characterized by the string
$$ T_4: v_1^{(1)}, v_2^{(1)};
v_4^{(1)}, v_3^{(1)};v_2^{(2)}, v_1^{(2)};v_3^{(2)}, v_4^{(2)}. $$
We express the parameter differences in terms of the weights,
$2\ell+1 = v^{(1)}-v^{(2)}$, and $\eps = v_2-v_3$.
We have the relations $ 2\ell_1+2\ell_3+2 = 0, \ 2\ell_2+2\ell_4+2 = 0 $.
\be \label{YSC4S12}
\bar \Phi =  x_{14}^{2\ell_1-2\ell_2 -1} x_{12}^{ 2\ell_2 +1}
x_{34}^{-2\ell_1-1} x_{23}^{-1} \left ( \frac{x_{23} x_{14} }{x_{12}
x_{34}}\right)^{\eps}. \ee 
The result does not differ essentially from  (\ref{4corr1}). 
It takes to change to the dual representation
in the points 2 and 4, $\ell_2 \to \bar \ell_2= -\ell_2-1$,
$2\bar \ell_2+2 \bar \ell_4 +2 = 0 $ and to do a cyclic shift in the point
labels $1,2,3,4 \to 2,3,4,1$ to transform $\bar \Phi$ into 
the ratio-coordinate factor in (\ref{4corr1}).

\subsection{ A 6-point correlator }

We construct a 6-point correlator appearing in subsection \ref{sec:5.2} as kernel of a generalised
Yang-Baxter operator. We consider the construction in homogeneous
coordinates only. The resulting monodromy and the relation between the
spectral parameters, the dilatation weights and two extra parameters 
will be  relevant later.

This correlator can be constructed by $R$ operator action on $ \Phi_0 =
\delta^{(2)}(\mathbf{x}_1) \delta^{(2)}(\mathbf{x}_2)$ $\delta^{(2)}(\mathbf{x}_3) $.
\begin{align*}
 \Phi_6 
& = R^{++}_{65}(u_3^+ - u_2^+) R^{++}_{54}(u_3^+- u_1^+) R^{++}_{43}(u_3^+ -u_6)
R^{++}_{65}(u_2^+ - u_1^+) R^{++}_{54}(u_2^+- u_6) \\ 
& \hspace{3.5cm} \times R^{++}_{42}(u_2^+ -u_5) 
R^{++}_{65}(u_1^+ - u_6) R^{++}_{54}(u_1^+- u_5) R^{++}_{41}(u_1^+ -u_4)
\Phi_0 \\
& = \int \mathrm{d}^9c \, \varphi_6(c) \, \prod_{i=1}^3 \delta^{(2)} (\mathbf{x}_i -
\sum_{j=4}^6 c_{ij} \mathbf{x}_j ). 
\end{align*}
The permutation pattern is 
$$ \begin{pmatrix}
 1  & 2  & 3  & 4  & 5  & 6 \\
 4  & 5  & 6  & 1  & 2  & 3
\end{pmatrix}.
$$
It allows to express the weights in terms of the 
spectral parameters,
$$ 2\ell_1 = u_4 - u_1, \ \ 2\ell_2 = u_5 -u_2, \ \ 2\ell_3 = u_6 - u_3. $$
The other weights are dependent by
$$ 2\ell_i + 2\ell_{i+3} + 2 = 0, \  i=1,2,3. $$
We express the parameters $u_1,u_2,u_3$ in terms of weights and then the remaining
ones  enter the result by the two independent parameters
$ \varepsilon_{4} = u_4 - u_5,  \varepsilon_{5} = u_5 - u_6  $.
The monodromy is  $(u= u_6)$
\begin{multline} \label{T6} T_6 = L^+_1(u_1)  L^+_2(u_2)  L^+_3(u_3)  L^+_4(u_4)  L^+_5(u_5) L^+_6(u_6) \\ 
= L^+_1(-2\ell_1 + \e_4 + \e_5 + u) L^+_2(-2\ell_2 +\e_5 + u) L^+_3(-2\ell_3 +u)
L^+_4(\e_4 + \e_5 +u) L^+_5(\e_5+u) L^+_6(u). \end{multline}
The inverse of the link variable function $\varphi_6(c)$ in the integrand is
expressed in terms of the consecutive minors of the rectangular matrix
$$ \begin{pmatrix}
 1  & 0  & 0  & -c_{14}  & -c_{15}  & -c_{16} \\
 0  & 1  & 0  & -c_{24}  & -c_{25}  & -c_{26} \\
 0  & 0  & 1  & -c_{34}  & -c_{35}  & -c_{36} 
\end{pmatrix}.
$$
denoted by triples of column numbers $(ijk)$.
This is a feature  pointed out in 
context of scattering amplitudes \cite{AH}.
\begin{align*} \varphi_6(c)^{-1} &= (2 3 4 )^{1+u_5 -u_4} (3 4 5 )^{1+u_6 -u_5} (4 5 6 )^{1+
u_1^+ -u_6} (5 6 1)^{1+ u_2^+ - u_1^+} (6 1 2)^{1+ u_3^+ -u_2^+} 
\\
&=(2 3 4 )^{1-\varepsilon_{4}} (3 4 5 )^{1-\varepsilon_{5}} 
(4 5 6 )^{1-2\ell_1 -2+ \varepsilon_{4}+\varepsilon_{5}} 
(5 6 1)^{1+ 2\ell_1 - 2\ell_2 -\varepsilon_{4}} 
(6 1 2)^{1+2\ell_2 - 2\ell_3 - \varepsilon_{5}}.
\end{align*}
In Appendix \ref{app:B} we do six out of the nine integrations by the $\delta$-distributions.
The result is
\be \label{6corr1}
  \Phi_6 =
 \langle 45 \rangle^{-1 - 2\ell_2 - \varepsilon_{4}} \,  \int \frac{\mathrm{d}c_{16} \mathrm{d}c_{26} \mathrm{d}c_{36}}
{ c_{36}^{1+2\ell_2 - 2\ell_3 -\varepsilon_{5}} } \, \phi_6, \ee \vspace{-20pt}
\begin{multline*} \phi_6^{-1} =
 (\langle 15 \rangle - c_{16} \langle 65 \rangle )^{1-\varepsilon_{4}} 
( \langle 12 \rangle - c_{16} \langle 62 \rangle + c_{26}\langle 61 \rangle )^{1 -\varepsilon_{5}}
  \\
  \times (\langle 12 \rangle c_{36} + \langle 23 \rangle c_{16} + \langle 31 \rangle c_{26} )^{-1-2\ell_1
+\varepsilon_{4} +\varepsilon_{5}} \ 
(\langle 24\rangle c_{36} - \langle 34 \rangle c_{26} )^{1+2\ell_1-2\ell_2  -\varepsilon_{4}}. 
\end{multline*}

\section{Correlators with deformed symmetry} \label{sec:4}
\setcounter{equation}{0}

The symmetry condition for correlators can be extended to the cases
with the monodromy matrix operator built from $L$ matrices with
algebraic deformations. In applications algebraic deformation may describe
particular ways of symmetry breaking. 

In the  case of trigonometric (q-) deformation 
and also for the case of elliptic deformation
we know the expressions for the $L$ matrices, 
 the YB operators and of its factors in analogy \cite{DKK07}.
We discuss the case of quantum deformation only.

The straightforward q-deformations of (\ref{L+-})
are
\be \label{Lq+-}
L_q^+(u) = \begin{pmatrix}
[u+1 +x_1\partial_1]_q & \frac{x_2}{x_1} [x_1\partial_1]_q \\ \frac{x_1}{x_2}
[x_2\partial_2]_q & [u+1 + x_2\partial_2]_q
\end{pmatrix},
\ \ 
L_q^-(u) =
\begin{pmatrix}
[u -x_1\partial_1]_q & -\frac{x_1}{x_2} [x_2\partial_2]_q 
\\ -\frac{x_2}{x_1} [x_1\partial_1]_q & [u- x_2\partial_2]_q
\end{pmatrix}.
\ee
We use the notation $[x]_q = (q^x - q^{-x}) (q-q^{-1})^{-1} $.
We have the factorized forms
\begin{multline*} L_q^+(u) = (q - q^{-1})^{-1} 
\begin{pmatrix}
1 &1 \\ -\frac{x_1}{x_2} q^{-u-1- x_1\partial_1 -x_2\partial_2} &
-\frac{x_1}{x_2} q^{u+1+x_1\partial_1 +x_2\partial_2}
\end{pmatrix}
\begin{pmatrix}
q^{x_1\partial_1+1} & 0 \\ 0 & q^{-x_1\partial_1 -1}
\end{pmatrix} \\
\times \begin{pmatrix}
q^{u} & \frac{x_2}{x_1} \\
-q^{-u} & -\frac{x_2}{x_1}  
\end{pmatrix},
\end{multline*}
$$L_q^-(u) = (q - q^{-1})^{-1}
 \begin{pmatrix}
1 &1 \\ \frac{x_2}{x_1} q^{-u } & \frac{x_2}{x_1} q^{u}
\end{pmatrix}
\begin{pmatrix}
q^{x_2\partial_2+1} & 0 \\ 0 & q^{-x_2\partial_2 -1}
\end{pmatrix}
\begin{pmatrix}
q^{u-1-x_1\partial_1-x_2\partial_2} & -\frac{x_1}{x_2} \\
-q^{-u+1+x_1\partial_1+x_2\partial_2} & \frac{x_1}{x_2}
\end{pmatrix}.
$$
With these forms it is easy to prove that the relation of inversion
(\ref{inv}) hold with a modification.
\be \label{invq}
 \left ( \frac{L_q^+ (u)}{[u]_q} \right )^{-1} = \frac{L_{q^{-1}}^+(-u-1-
(x_1 \dd_1 + x_2 \dd_2))}{
[-u-1- (x_1 \dd_1 + x_2 \dd_2)]_q}, 
\ee
$$
\left ( \frac{L_q^- (u)}{[u]_q} \right )^{-1} = \frac{L_{q^{-1}}^-(-u+1+
(x_1 \dd_1 + x_2 \dd_2))}{
[-u +1+ (x_1 \dd_1 + x_2 \dd_2)]_q}.   
$$
Here $x_1,x_2$ are the components of a 2-dimensional vector
$\mathbf{x}$. 
In connection with correlators and YB relation we deal with
more than one 2-dimensional vectors and we use the index notation
$x_{i,1}, x_{i,2}$ for the components of $\mathbf{x}_i$. 
As in the undeformed case
we introduce the ratio variables for the point $i$ if
 the corresponding $L$ matrix is chosen with signature
$+$ as $ x_i = \frac{x_{i,1}}{x_{i,2}} $
or with signature $-$  as
$x_i = - \frac{x_{i,2}}{x_{i,1}} $.
Then we have for the restriction to the weight $2\ell$
in ratio coordinates
\begin{multline} \label{L+r}
L_{q}^+(u^+, u) \to L_q(u^++1,u) = 
\begin{pmatrix}
[x\partial+u+1]_q & x^{-1}[x\partial]_q \\
x [-x\partial + u^+-u]_q & [u^+ +1 -x\partial]_q
\end{pmatrix} \\
=[u^+ +1]_q \hat V^{-1}_q(u^+ +1) \hat D_q \hat V_q(u), 
\end{multline}
$$
\hat{V}_q(u) = \begin{pmatrix}
q^u & x^{-1} \\ - q^{-u} & -x^{-1}
\end{pmatrix}, \ \ \ 
\hat D_q =
\begin{pmatrix}
q^{x\partial +1} & 0 \\ 0 & q^{-x\partial-1}
\end{pmatrix},
$$
and similar for $L^-(u,u^-)$ in complete analogy to the undeformed case
(\ref{Lratio}).

We introduce also $v^{(1)}, v^{(2)}$ as above (\ref{v1v2}) and write the inversion
relation in analogy,
\be \label{invqv12}
L_q^{-1 }(v^{(1)}, v^{(2)})= - \frac{1}{[v^{(1)}]_q \ [v^{(2)}]_q}
L_{q^{-1}} (-v^{(2)},-v^{(1)}).
\ee
The YB operator factors corresponding to the elementary permutations
of the parameters $v_1^{(1)}, v_1^{(2)}, v_2^{(1)}, v_2^{(2)} $
in the product of $L$ operators are known \cite{DKK07}.
\be \label{S12q}
 \mathcal{S}_{q12}(w) = x_1^{w} \frac{(\frac{x_2}{x_1} q^{1-w};
q^2)}{(\frac{x_2}{x_1} q^{1+w}; q^2)} = \pi(x_1,x_2;w), \ \ w= v_1^{(2)}-
v_2^{(1)}. \ee
The latter notation emphasizes that this is the appropriate deformation of
$(x_1-x_2)^w $. We use the standard notation for the infinite product
$(x;q):= \prod_{j=1}^\infty (1-xq^{j-1})$.

The other YB operators of elementary permutations are in analogy
given in terms of the intertwiner $W_q$, 
$\mathcal{S}_{q11}(v_1^{(1)}- v_1^{(2)}) = W_{q, 1}(v_1^{(1)}- v_1^{(2)})$, $
\mathcal{S}_{q22}(v_2^{(1)}- v_2^{(2)}) = W_{q, 2} (v_2^{(1)}- v_2^{(2)}) $.
\be \label{intertWq}
 W_{q, i}(a)= \frac{1}{x_i^a} \frac{( q^{2x_i\dd_i +1-a};
q^2)}{( q^{2x_i\dd_i+2}; q^2)} \ q^{-a x_i\dd_i} =
\frac{1}{x_i^a} \frac{ \Gamma_q (x_i\dd_i +1)}{ \Gamma_q (x_i\dd_i+1-a)}. \ee
The latter notation emphasizes that the expression  can be considered 
as the deformation of (\ref{intertW}) with the ratio of Gamma functions.
The analogy is useful indeed because there is a q-deformed analogy of the
Beta integral \cite{KKM02} allowing to express the  action of $W_q$ on functions
in analogy to (\ref{Wf}) as
\be \label{Wfq}
  W_q(a) f(x) = \mathrm{const} \int \mathrm{d}\tilde x \, \pi(x,\tilde x; -a-1) f(\tilde x). \ee

The YB relation for $\mathcal{S}_{q 12}$ can be checked directly 
using the factorized form (\ref{L+r}) \cite{DKK07}, see Appendix \ref{app:C}.
\begin{equation} \label{eq:RLL-rational}
\mathcal{S}_{q 12 } (v_1^{(2)}-v_2^{(1)}) L_1(v_1^{(1)},v_1^{(2)})
L_2(v_2^{(1)},v_2^{(2)}) =  
L_1(v_1^{(1)},v_2^{(1)}) L_2(v_1^{(2)},v_2^{(2)}) \mathcal{S}_{q12
}(v_1^{(2)}-v_2^{(1)}).
\end{equation}

From  the deformed $\mathcal{S}_{q 12}(u)$ it is straightforward to reconstruct
the deformed $R^{+-}_{q 12 }$,
\begin{equation}
R^{+-}_{q 12 }(w)  = (x_{1,1}x_{2,1})^w \cdot 
\frac{(\mathcal{X} q^{1-w};
q^2)}{(\mathcal{X} q^{1+w}; q^2)}, \ \  \
\mathcal{X}=-\frac{x_{1,2}x_{2,2}}{x_{1,1}x_{2,1}}.
\end{equation}
The factorized form in homogeneous coordinates allows to repeat the above
calculation and to check the YB relation (see Appendix \ref{app:C})
\begin{equation} \label{RqLL}
R^{+-}_{q 12}(u_1-u_2) L^+_{q 1}(u_1) L_{q 2}^-(u_2) =  L_{q 1}^+(u_2) L_{q
2}^-(u_1) R_{q 12}(u_1-u_2).
\end{equation}

The examples of symmetric correlator constructions working in ratio
coordinates and the YB operators of elementary permutations work in the
deformed case by straightforward analogy. The basic correlator $1$ from which we generate the result by $R$ operations
has vanishing  weights. The resulting correlator has
weights calculated from the permutation of parameters to
$ v_1^{(1)}, v_1^{(2)}; v_2^{(1)}, v_2^{(2)}; v_3^{(1)},v_3^{(2)}; \dots$ as
$ 2\ell_i +1 =  v_i^{(1)}- v_i^{(2)} $.

The action of (\ref{eq:RLL-rational}) on $1$ shows that 
$$ \mathcal{S}_{q12}(w) = \pi(x_1,x_2;w) =    x_1^w 
\frac{(\frac{x_2}{x_1}q^{1-w},q^2)}{(\frac{x_2}{x_1}q^{1+w},q^2)}, \ \ \ 
T_2:v_1, v_2; v_1-1,v_2-1,
$$ with 
$w= v_1-v_2-1$ is the symmetric two-point correlator for the monodromy
$T_2$.

Let us mention an example of a 3-point correlator.
We have
\begin{multline*}
\mathcal{S}_{q23}(v_2^{(2)}-v_3^{(1)}) \mathcal{S}_{q12}(v_1^{(2)}-v_2^{(1)}) 
L_1(v_1^{(1)}, v_1^{(2)}) L_2(v_2^{(1)}, v_2^{(2)}) 
L_3(v_3^{(1)}, v_3^{(2)}) \\
= L_1(v_1^{(1)}, v_2^{(1)}) L_2(v_1^{(2)}, v_3^{(1)}) L_3(v_2^{(2)}, v_3^{(2)}) 
\mathcal{S}_{q23}(v_2^{(2)}-v_3^{(1)}) \mathcal{S}_{q12}(v_1^{(2)}-v_2^{(1)}).
\end{multline*}
Applying both sides on the constant function $1$ we obtain
the symmetric 3-point correlator for the monodromy 
$T_3:v_1, v_2; v_1-1, v_3; v_2-1,v_3-1$ as 
$$ \mathcal{S}_{q23}(v_2^{(2)}-v_3^{(1)}) \mathcal{S}_{q12}(v_1^{(2)}-v_2^{(1)}) = 
x_1^{v_1-v_2-1} x_2^{v_2-v_3-1} 
\frac{(\frac{x_3}{x_2}q^{2+v_3-v_2};q^2) }{(\frac{x_3}{x_2}q^{v_2-v_3};q^2)} 
\frac{(\frac{x_2}{x_1} q^{2+v_2-v_1};q^2) }{(\frac{x_2}{x_1}q^{v_1-v_2};q^2)},
$$
The eigenvalue is 
$ E(v) = [v_1]_q [v_2]_q [v_3]_q$.

The result of the 3-point correlator 
(\ref{YSCS12}) and the ones of the two steps of its construction
lead to the corresponding deformed correlators with the monodromy
characterized by the same parameter strings with the substitutions
$$ (x_i-x_j)^w \rightarrow \pi(x_i,x_j;w). $$  
This substitution rule works for all YSC which can be generated 
by $\mathcal{S}_{ij}$ actions ($i\not=j$) in this way, in particular for the
4-point YSC (\ref{YSC4S12}).

\section{Integral operators from symmetric correlators } \label{sec:5}
\setcounter{equation}{0}

We consider integral operators with YSC as kernels. The  symmetry
of the correlators expressed in the monodromy relation (\ref{YSC})
implies  the symmetry of the operators by the relation of inversion and
transposition obeyed by the involved $L$ operators.

\subsection{Generalized Yang-Baxter operators} \label{sec:5.1}

The YB  operators considered above map the tensor product of
representation spaces as $V_{\ell_1} \otimes V_{\ell_2} \to V_{\tilde \ell_1}
\otimes V_{\tilde \ell_2} $ and obey a relation with a product of two $L$
matrices of the form
$$ \mathrm{R}\, L^+_1(u_1^+, u_1) L^+_2(u_2^+, u_2) = 
L^+_{\sigma (1)}(u_{\sigma (1)}^{\prime +}, u_{\sigma (1)}\p) L^+_{\sigma
(2)}(u_{\sigma (2)}^{\prime +}, u_{\sigma (2)}\p ) \, \mathrm{R}. $$
We consider generalized $R$ operators mapping tensor products with more
factors and obeying the more general relation involving monodromy matrices
of $M$ factors. 
\be \label{genRLL}
 \mathrm{R}\, L^+_1(u_1^+, u_1)\cdots L^+_M(u_M^+, u_M) = 
L^+_{\sigma (1)}(u_{\sigma (1)}^{\prime +}, u_{\sigma (1)}\p)\cdots L^+_{\sigma
(M)}(u_{\sigma (M)}^{\prime +}, u_{\sigma (M)}\p) \, \mathrm{R}. \ee
The interchange results in another monodromy matrix with the $L$ factors
permuted, $1, \dots,M \to \sigma(1), \dots,\sigma(M)$ and the sets of parameters
substituted $u_1^+ ,\dots, u_M^+ \to u_1^{\prime +}, \dots,$ $ u_M^{\prime +}$,  and
 $u_1 ,\dots, u_M \to u_1\p,\dots, u_M\p$. 

Examples of such operators are obtained by constructing integral operators
with the kernels chosen as particular Yangian symmetric correlators.
The Yangian symmetry condition then implies a generalized YB relation.
Indeed, let the operator acting on a function of $M$ points be defined as
\be \label{intR}
 \mathrm{R} \psi(\mathbf{x}_1, \dots, \mathbf{x}_M) = 
\int \mathrm{d}\mathbf{x}_1\p  \cdots \mathrm{d}\mathbf{x}_M\p
\psi(\mathbf{x}_1\p, \dots, \mathbf{x}_M\p) \Phi(\mathbf{x}_1\p, \dots,
\mathbf{x}_M\p, \mathbf{x}_1, \dots, \mathbf{x}_M). \ee
Both the function and the kernel depend homogeneously on the coordinates
of the points with weights $2\tilde \ell_1, \dots,2\tilde \ell_M$ for the
function  and $2 \ell_1\p, \dots,2 \ell_M\p, 2\ell_1, \dots,2 \ell_M$ for the
kernel with  the relations
$$  \ 2\ell_i\p + 2\tilde \ell_i +2= 0, \quad i=1, \dots,M. $$ 
The resulting function has the weights $2\ell_1, \dots,2 \ell_M$, and the case
$2\ell_i = 2\tilde \ell_i$ will be considered in particular.

The Yangian symmetry condition obeyed by the kernel $\Phi$ has the general
form (\ref{YSC}) and can be rewritten after relabeling
$1, \dots,N \to 1\p, \dots,M\p, 1, \dots,M$ as
$$L^+_1(u^+_1, u_1)\cdots L^+_M(u^+_M,u_M) \Phi = E(\mathbf{u}) 
L_{M\p}^{+ -1}(u^+_{M\p}, u_{M\p}) \cdots L^{+ -1}_{1\p}
(u^+_{1\p},u_{1\p}) \Phi. $$
We consider the action of $L^+_1(u^+_1, u_1)\cdots L^+_M(u^+_M, u_M)$ onto
$\mathrm{R} \psi$,  apply the latter relation for the kernel and
move by transposition the action of $L_{M\p}^{+ -1}(u^+_{M\p}, u_{M\p})
\cdots L^{+ -1}_{1\p}(u^+_{1\p},u_{1\p})$
from the kernel to the function.
\begin{align*} 
& L^+_1(u^+_1, u_1) \cdots L^+_M(u^+_M, u_M)\, \mathrm{R}\, \psi(\mathbf{x}_1,\dots , \mathbf{x}_M) \\
& \quad= \int d\mathbf{x}_1\p  \cdots \mathbf{x}_M\p
\psi(\mathbf{x}_1\p,\dots, \mathbf{x}_M\p) \cdot
L^+_1(u^+_1, u_1) \cdots L^+_M(u^+_M, u_M) \Phi(\mathbf{x}_1\p, \dots,\mathbf{x}_M\p, 
\mathbf{x}_1,\dots, \mathbf{x}_M) 
 \\
& \quad= E(\mathbf{u})  \int d\mathbf{x}_1\p \cdots \mathbf{x}_M\p
\left[L_{M\p}^{+ -1 T}(u^+_{M\p}, u_{M\p})\cdots L^{+ -1 T}_{1\p}(u^+_{1\p},u_{1\p}) \,
\psi(\mathbf{x}_1\p, \dots,\mathbf{x}_M\p)\right] \\
& \hspace{9.8cm} \times\Phi(\mathbf{x}_1\p, \dots,\mathbf{x}_M\p, \mathbf{x}_1,\dots,\mathbf{x}_M).  \end{align*}
We have the relation for inversion and operator conjugation of the $L$
matrices
\be \label{L-1T}
(L^{+ } (u^+, u) )^{-1 T} = \frac{1 }{u(u^+ +1) } L^+(u-2, u^+).
\ee
We abbreviate 
$$ F(\mathbf{u}) = \frac{E(\mathbf{u})}{u_{1\p}(u_{1\p}^{ +} +1) \cdots u_{M\p}
(u_{M\p}^{+} +1)}. $$
Then we obtain
$$  L^+_1(u^+_1, u_1) \cdots L^+_M(u^+_M, u_M)\, \mathrm{R}\, \psi = 
F(\mathbf{u})\, \mathrm{R} \,
L^+_M(u_{M\p} - 2, u_{M\p}^{ +}) \cdots L^+_1(u_{1\p} - 2, u_{1\p}^{ +})
\, \psi.  
$$
The result is of the form of (\ref{genRLL}) with $\sigma (1,2, \dots M) = M, M-1,
\dots,1$ and $u_{i\p} \to u_{i\p}^{ +}  $, $ u_{i\p}^{ +} \to u_{i\p}-2$.  

Integral operators with YSC kernels act in a highly symmetric way, defining
homomorphisms not only of the $s\ell_2$ Lie algebra action on the functions 
(global symmetry) but of the
related Yangian  algebra. The resulting YB relations   
encode  this symmetry property. It is also reflected in the
eigenvalue spectrum.

%%%%%%%%%%%%%%%%%%%%%%%%%%%%%%%%%%%%%%%%%%%%%%%%%%%%%%%%%%%%%%%%%%%%%%%%%%
\subsection{Examples} \label{sec:5.2}

We specify the construction of Yangian symmetric operators from YSC in the
cases $M=2$ and $M=3$. In the first case we obtain operators with simple 
commutation relations with a monodromy out of two $L$ operators, in
particular the standard $RLL$ Yang-Baxter relation. We use as kernels the 
YSC from the stepwise construction (\ref{Delta}, \ref{kernel}).
In the case $M=3$ we use the 6-point YSC (\ref{6corr1}) and the related
monodromy (\ref{T6}) to obtain the generalized YB relation involving the
monodromy out of three $L$ operators.   

\vspace{0.5cm}

{\it Symmetric two-point operators} \\
\noindent
We consider first integral operators with  4-point YSC as  kernels obeying
\be \label{Ys}
 T_{1,2,3,4}(u) \Phi^{--++} = E(u) \Phi^{--++} \ee
with the monodromy
$$  T_{1,2,3,4}(u) = L^+_1(u_1^+, u_1) L^+_2(u_2^+, u_2)
L^+_3(u^+_3, u_3) L^+_4(u^+_4, u_4) $$
resulting from the original monodromy 
$L^+_1(u^{0+}_1, u_1^0) \cdots L^+_4(u_4^{0+}, u^0_4) $ if 
the correlator is  generated
by $R^{++}$ operations from the basic correlator 
$ \delta^{(2)}(\mathbf{x}_1)  \delta^{(2)}(\mathbf{x}_2) $.
We denote those correlators by superscript $--++$.
Parameters $u_1, \dots, u_4$ and $ u^{+}_1, \dots, u^{ +}_4 $ are permutations of
$u^0_1, \dots,u^0_4$ and $u^{0 +}_1 = u^0_1-2, u^{0 +}_2 = u^0_2-2, 
u^{0 +}_3 = u^0_3, u^{0 +}_4 = u^0_4$. 
The eigenvalue function is $E(u) = u^0_{1} u^0_{2} (u^0_3 +1) (u^0_4+1) $.

We consider now functions $\psi(\mathbf{x}_1,\mathbf{x}_2) $ with the weights
$2\tilde \ell_1$ and $ 2\tilde \ell_2$ at the two points and the operator
defined by the specification of (\ref{genRLL})
\be \label{defQ}
 \hat Q \psi(\mathbf{x}_1,\mathbf{x}_2) = \int \mathrm{d}\mathbf{x}_{1\p} \mathrm{d}\mathbf{x}_{2\p}
\psi(\mathbf{x}_{1\p},\mathbf{x}_{2\p})
\Phi^{--++}(\mathbf{x}_{1\p},\mathbf{x}_{2\p}, \mathbf{x}_{1},\mathbf{x}_{2} ). \ee

Recall that at this step the initial labels of the points have been changed
as $1,2,3,4 \to 1\p, 2\p,1,2 $.  The scale symmetry of the integration
requires for the weights at the corresponding correlator points
$$ 2\ell_{i\p} = -2-2\tilde \ell_i. $$

Consider the action of $ L^+_{1}(u^+_1, u_1) L^+_2(u^+_2, u_2)$ on
$ \hat Q \psi(\mathbf{x}_1, \mathbf{x}_2) $
 and  rewrite the  symmetry condition (\ref{Ys}) in the redefined labels as
$$ L^+_1(u^+_1, u_1) L^+_2(u^+_2, u_2) \Phi^{--++}(x_{1\p},x_{2\p}, x_{1},x_{2} ) =
E(u) L_{2\p}^{+ -1}(u^+_{2\p}, u_{2\p}) L_{1\p}^{+ -1}(u^+_{1\p}, u_{1\p}) \Phi^{--++} $$
and conclude as above
\be \label{QLL}
 \hat Q L^+_2(u_{2\p}-2, u_{2\p}^+) L^+_1(u_{1\p}-2, u_{1\p}^+) F(u)
= L^+_1(u_1^+, u_1) L^+_2(u_2^+, u_2) \hat Q,
\ee\vspace{-10pt}
$$ F(u) = \frac{u^0_{1\p} u^0_{2\p} (u^0_1+1) (u^0_2 +1)}
{u_{1\p}( u^{ +}_{1\p}+1)u_{2\p} (u_{2\p}^+ +1) }.$$
The cases of monodromies are specified by the permutation pattern 
$$
\begin{pmatrix}
\sigma(1\p) & \sigma(2\p) & \sigma(1) & \sigma(2) \\
 \bar \sigma(1\p) &\bar \sigma(2\p) &\bar \sigma(1)&\bar \sigma(2)
\end{pmatrix}
$$
meaning
in particular the relation between the original parameters 
$ u_i^0, u_i^{0 +}$ and the
ones appearing in (\ref{Ys}), (\ref{QLL}),
$ u_i = u^0_{\sigma(i)}$, $ u^+_i = u^{0 +}_{\bar \sigma(i)}$.
In the following we shall formulate this as substitution rules for the
parameters in (\ref{QLL}) in terms of the original ones and omit the
superscript $0$ after substitution.

\vspace{.5cm}

{\it Point permutation operator}\\
\noindent
In the case of $\Delta^{--++} $ (\ref{Delta}) we have
$$ \begin{pmatrix}
1\p&2\p&1&2 \\
2&1&2\p&1\p   
\end{pmatrix}. $$
 This means
$$ u_{1\p} \to  u_{1\p}, u_{2\p} \to u_{2\p}, u_1 \to u_1, u_2 \to u_2 $$
$$ u_{1\p}^{ +} \to u_2, u_{2\p}^{ +} \to u_1,  
u_1^{ +} \to u_{2\p} -2, u_2^{ +} \to u_{1\p}-2 $$
and  implies $F(u) = 1$. We obtain as the specification of (\ref{QLL})
that the operator $\hat \Delta$
defined with this kernel obeys 
$$ \hat \Delta L^+_2(u_{2\p}-2, u_1) L^+_1(u_{1\p}-2, u_2) 
 = L^+_1(u_{2\p}-2, u_1) L^+_2(u_{1\p}-2, u_2) \hat \Delta.
$$
Thus choosing $\Delta^{--++} $ as the kernel in (\ref{defQ})
one  represents the operator of permutation $P_{12}$ acting as
$$ P_{12} (\mathbf{x}_1, \mathbf{p}_1) P_{12} = (\mathbf{x}_2,
\mathbf{p}_2).$$

\vspace{.5cm}

{\it Parameter pair permutation operators}\\ 
\noindent
The kernel $R^{++}_{21} \Delta^{--++} $ defines by (\ref{defQ})
the operator $\hat R^{++}_{12} = P_{12} R^{++}_{12}$. 
Indeed, the permutation pattern after
relabeling is
$$ \begin{pmatrix}
1\p&2\p&1&2 \\
1&2&2\p&1\p
\end{pmatrix}. $$
 This means
$$ u_{1\p} \to u_{1\p}, u_{2\p} \to u_{2\p}, u_1 \to u_1, u_2 \to u_2 $$
$$ u_{1\p}^{ +} \to u_1, u_{2\p}^{ +} \to u_2,  
u_1^{ +} \to u_{2\p} -2, u_2^{ +} \to u_{1\p}-2. $$
We see that $F(u) = 1$ and the relation (\ref{QLL}) reads
$$\hat R_{12}^{++} L^+_2 (u_{2\p} -2, u_2)  L^+_1(u_{1\p} -2, u_1)
= L^+_1(u_{2\p} -2, u_1) L^+_2(u_{1\p} -2, u_2) \hat R_{12}^{++}. 
 $$
The parameters $u_1, u_2$ and the underlying canonical pairs are permuted.

The kernel $R^{++}_{12} \Delta^{--++} $ defines by (\ref{defQ})
the operator $\hat R^{++}_{21} = P_{12} R^{++}_{21}$. 
Indeed, the permutation pattern after
relabeling is
$$ \begin{pmatrix}
2\p&1\p&1&2 \\
2&1&2\p&1\p
\end{pmatrix}. $$
 This means
$$ u_{1\p} \to u_{2\p}, u_{2\p} \to u_{1\p}, u_1 \to u_1, u_2\to u_2, $$
$$ u_{1\p}^{ +} \to u_2, u_{2\p}^{ +} \to u_1, 
u_1^{ +} \to u\p_{2\p}-2, u_{2}^{ +} \to u_{1\p} -2.
$$
Again $F(u) = 1$. The relation (\ref{QLL}) results in
$$ \hat R_{21}^{++} L_2 (u_{1\p} -2, u_1)  L_1(u_{2\p} -2, u_2)
=L_1(u_{2\p} -2, u_1) L_2(u_{1\p} -2, u_2) \hat R_{21}^{++}. 
 $$
The parameters $\tilde u_1^+ = u_{2\p}-2, \tilde u_2^+ = u_{1\p}-2$ are
permuted together with the underlying canonical pairs.

\vspace{.5cm}

{\it The complete Yang-Baxter operator}  \\ 
\noindent
The kernel $ R^{++}_{12} R^{++}_{21} \Delta^{--++}= \Phi^{--++}$ (\ref{kernel})
defines by (\ref{defQ})
the  complete YB operator $\hat \R_{12}$. 
Indeed, the permutation pattern after
relabeling is   
$$ \begin{pmatrix}
2\p&1\p&1&2 \\
1&2&2\p&1\p
\end{pmatrix}. $$
 This means
$$u_{1\p} \to u_{2\p}, u_{2\p} \to u_{1\p}, u_1 \to u_1, u_2 \to u_2, $$
$$ u_{1\p}^{ +} \to u_1, u_{2\p}^{+} \to u_2, 
u_1^{ +} \to u_{2\p}-2, u_{2}^{ +} \to u_{1\p} -2.
$$
We find $F(u) = 1$ and the relation (\ref{QLL}) results in
\be \label{YBnormal}
 \hat \R_{12} (u_2-u_1) L^+_2(u_{1\p} -2, u_2)  L^+_1(u_{2\p} -2, u_1)
= L^+_1(u_{2\p} -2, u_1) L^+_2(u_{1\p} -2, u_2) \hat \R_{12}(u_2-u_1). 
 \ee
Both parameters and the  underlying canonical pairs are permuted.
The weights are $2\ell_1 = u_{2\p} - u_1-2$, $2\ell_2 = u_{1\p} - u_2 -2$ and
after fixing their values one can omit the $u^+$ arguments in $L_i^+ (u_i^+,
u_i)$  to match the conventional form. 
The explicit form of the kernel is given by (\ref{kernel}) with the 
substitution $1,2,3,4 \to 1\p, 2\p, 1,2$.

The completely connected 4-point correlator has been constructed by other 
$R$ sequences, e.g. 
string with the correlator (\ref{4corr1}). Corresponding to the permutation
pattern
$$ \begin{pmatrix}
1\p&2\p&1&2 \\
1&2&1\p&2\p   
\end{pmatrix} $$
one obtains again the normal YB relation (\ref{YBnormal})
with other parameter substitutions and with the weights  calculated differently 
from the original spectral parameters.

Note that the permutation $P_{12}$ can be removed by modifying the
definition (\ref{defQ}) of the integral operator by a permutation of $1,2$ or 
$1\p, 2\p$ in the step of relabeling the original $1,2,3,4$.

\vspace{.5cm}

{\it Explicit kernel in ratio coordinates} \\ 
\noindent
The YB relation with $\hat \R_{12}(u_1-u_2)$ (\ref{YBnormal})
can be written in the 
ratio coordinate form (sect. \ref{sec:2.3})
with $\hat{ \mathrm{R}}_{12}(v_1-v_2) = P_{12} \mathrm{R}^1_{12} \mathrm{R}^2_{12}$
\be \label{RLLv}
  \hat{ \mathrm{R}}_{12}(v_1- v_2) 
L_1(v_1^{(1)},v_1^{(2)}) L_2(v_2^{(1)},v_2^{(2)})  =
L_2(v_2^{(1)},v_2^{(2)}) L_1(v_1^{(1)},v_1^{(2)})
\hat{ \mathrm{R}}_{12}(v_1-v_2). \ee
The relation of reduction of $L^+(u)$  to $L(v^{(1)},v^{(2)})$ (\ref{Lred})
implies that
the factor in  ratio coordinates in the 4-point correlator (\ref{kernel})
$$  \frac{x_{12}^{1+ 2 \ell_1 }}{x_{14} 
x_{2 3}^{1+ 2\ell_1-2\ell_2} x_{34}^{1+2\ell_2} }
\left ( \frac{x_{1 2}  x_{34} }{ x_{14} x_{23} } \right )^{u_4-u_3}
$$
results in the kernel of $P_{12} \mathrm{R}_{12}$.
Our
convention about the notations $u$ and $v$ is $v_i = u_i + \ell_i$, i.e.
$u_4 -u_3 = v_4 -v_3+\ell_3 - \ell_4 $. We recall that the weights of the
correlator (\ref{kernel}) are related as $2\ell_1 + 2\ell_3 +2 = 0$, $2\ell_2
+ 2 \ell_4+ 2= 0 $. According to (\ref{defQ}) the kernel of $P_{12}
\mathrm{R}_{12}$ is obtained from the resulting expression by relabeling
the points as $1, 2, 3, 4 \to 1\p, 2\p,1, 2$. Finally, we obtain the
kernel of $ \mathrm{R}_{12}(v), v= v_2-v_1$, 
(without the permutation $P_{12}$)  by permuting  $1\p, 2\p$.  
\be \label{kernelRv}
 \tilde \Phi^{--++} =  
 \frac{x_{1 2}^{1+  \ell_1+\ell_2 }}{x_{11\p}^{1+ \ell_2-\ell_1} 
x_{2 2\p}^{1+ \ell_1-\ell_2} x_{1\p 2\p }^{1+\ell_1+\ell_2} }
\left ( \frac{x_{1 2}  x_{1\p 2\p} }{ x_{11\p} x_{22\p} } \right )^{v}. 
\ee
This result can be derived directly from the $RLL$ relation (\ref{RLLv}) 
using the factorized form of $L(v^{(1)},v^{(2)})$
(\ref{Lratio}), see e.g. \cite{DKK01}.

\vspace{.5cm}

{\it A three-point generalized Yang-Baxter operator} \\ 
\noindent
We have constructed the 6-point correlator $\Phi_6$ (\ref{6corr1})
with the weight relations with the monodromy (\ref{T6}).
The correlator has been  derived from the condition
\be \label{Mew6}
 T_6 \Phi_6 = E(u) \Phi_6,
\ee \vspace{-15pt} $$ E(u) = u_1 u_2 u_3 (u_4+1) (u_5+1)(u_6+1). $$ 
We relabel the points as $1,2,3,4,5,6 \to 1\p,2\p,  3\p,1,2,3$ and
consider the integral operator acting as
\be \label{defQ123}
 \hat Q \, \psi (\mathbf{x}_1,\mathbf{x}_2,\mathbf{x}_3)    = 
\int \mathrm{d} \mathbf{x}_1\p \mathrm{d} \mathbf{x}_2\p \mathrm{d} \mathbf{x}_3\p
\, \psi(\mathbf{x}_1\p,\mathbf{x}_2\p,\mathbf{x}_3\p) \,
\Phi_6(\mathbf{x}_1\p,\mathbf{x}_2\p,\mathbf{x}_3\p,\mathbf{x}_1,
\mathbf{x}_2,\mathbf{x}_3). \ee
Following the general scheme 
we act on $ \hat Q\psi$  (in the new labels) by 
$ L^+_1(\e_4 + \e_5 +u) L^+_2(\e_5+u) L^+_3(u) $. We use the
monodromy eigenvalue relation for $\Phi_6$ (\ref{Mew6}) rewritten by
 applying  the inversion and transposition relation (\ref{L-1T}). $E(u)$ cancels
and we obtain that $\hat Q = \R_{321}$ obeys the following generalized YB
relation (we omit the arguments $u^+_i = u_i +2\ell_i$, $i= 1,2,3 $)
\be \label{genYB3}
\hat  \R_{321}(\eps_4,\eps_5) L^+_3(u) L^+_2(\e_5+u) L^+_1(\e_4 + \e_5 +u)    = 
L^+_1(\e_4 + \e_5 +u) L^+_2(\e_5+u) L^+_3(u) \hat \R_{321}(\eps_4,\eps_5).    \ee

%%%%%%%%%%%%%%%%%%%%%%%%%%%%%%%%%%%%%%%%%%%%%%%%%%%%%%%%%%%%%%%%%%%%%
%%------------------------------------------------------------------
%%%%%%%%%%%%%%%%%%%%%%%%%%%%%%%%%%%%%%%%%%%%%%%%%%%%%%%%%%%%%%%%%%%

\subsection{Kernels of  QCD parton  evolution}

The generalized parton distributions  \cite{SPD} contribute to hard-exclusive
production processes. The form in (one-dimensional, light-cone component) momenta
is normally used. The form in positions (on the light ray) is convenient for
representing the conformal symmetry actions and is useful also in the related
operator product renormalization.  For explaining this schematically (for
more details cf. \cite{DKM13})
let us consider composite operators of the form
$$ D_{\mu_1} ...  D_{\mu_n} \psi(0) \cdot D_{\nu_1} ...  D_{\nu_m} \phi (0) 
$$
By contracting the indices with parallel light-like vectors
$x_1^{\mu}, x_2^{\nu}, x_{1,\mu} x_1^{\mu} = x_{2,\mu} x_2^{\mu}
= x_{1,\mu} x_2^{\mu}= 0$ one projects on the symmetric traceless part.
Let us choose the axial gauge $x_1^{\mu} A_{\mu} = 0$, then
$x^{\mu} D_{\mu} = x \dd$. It is natural to  consider  the generating
function of such operator product projections
$$ \sum_{n,m} \frac{x_1^n}{n!} \dd^n \psi(0) \cdot \frac{x_2^m}{m!} \dd^m
\phi(0) = \psi(x_1) \phi(x_2) $$
The resulting operator valued function of two points on the light ray can be 
substituted  instead of $\psi(x_1,x_2)$ in the definition of the integral
operator  (\ref{defQ}). Operator products involving
more than two fields can be treated analogously. Then we deal with
functions of $M$ points on the light ray and integral operators with $2M$
point YSC as kernels. Choosing the kernel appropriately, this results in a
convenient formulation of the renormalization of a class of composite
operators. 

In QCD, if choosing the light-cone components of the quark and gluon
field, these composite operators are called quasi-partonic \cite{Levrep}. 
In $\mathcal{N} = 4$ super Yang Mills theory one has   
analogous composite operators with  components of the superfield. 
With combinations of the scalar components one has  
in particular the  $SL(2)$ sector \cite{Beisert03} such that 
the  mixing with the products involving other components is absent.  
The spin chain treatment of the latter has attracted much attention because
of the non-compact representations at the sites.

The ordinary parton distributions \cite{DGLAP} contribute to the
(inclusive) deep-inelastic scattering structure function. Their evolution
kernels correspond to the vanishing momentum transfer limit $k_1+k_2=0$  
of the  kernels of generalized parton distributions.

We consider the kernel of the YB operator $\mathrm{R}_{12}(v)$ in ratio
coordinate form (\ref{kernelRv}). For real values of the positions and if
$x_{1\p}, x_{2\p} $ lie in between $x_1, x_2$ we can write this in the
simplex integral form. We consider the case $\ell_1= \ell_2 = \ell$. 
 $$
\tilde \Phi^{--++} =
\int_0^1 \mathrm{d}^3\alpha\, \delta(\alpha_1+\alpha_2+\alpha_3 -1) 
\alpha_1^{-1-v} \alpha_2^{-1-v}\alpha_3^{-1+v -2\ell} \ 
\delta(x_{11\p} - \alpha_1 x_{12}) \delta (x_{22\p} + \alpha_2 x_{12}) 
 $$ $$
\equiv J_{-v,-v,v-2\ell}.$$ 
This form is the convenient one for comparison with the evolution kernels of
generalized parton distribution as given in \cite{DK01}. It is also convenient
because the Fourier transformation is easily done. The resulting form
is the one of \cite{BKL} and $J_{n_1,n_2,n_3}$ is the notation for the terms
appearing in the kernels introduced in that papers.

In the limit $v\to 0$ and in the negative integer values for $v= -1, -2$ we 
find the particular cases related to the parton kernels \cite{DK01}.

At $v\to 0$ we find in the residues at $\alpha_1= 0$ and $\alpha_2= 0$
the kernel of the unit operator $\delta(x_{11\p}) \delta(x_{22\p}) $ and
also  
$$ J_{11\p}^{(-2h)} + J_{22\p}^{(-2h)}, \ \ \
    J_{11\p}^{(-2h)} = \int \mathrm{d}\alpha \frac{(1-\alpha)^{-2h}}{[\alpha]_+}
\delta(x_{11\p})  \delta(x_{22\p} + \alpha x_{12}). $$

At $v=-1$ we have from (\ref{kernelRv}) the  kernel
$$ J_{1,1, -1-2\ell} = \left(\frac{x_{1\p2\p}}{x_{12}} \right)^{-2-2\ell} \ x_{12}^{-2}. 
 $$
At $v= -2$ we have  (\ref{kernelRv})
$$ J_{2,2,-2-2\ell} = \left(\frac{x_{1\p2\p}}{x_{12}} \right)^{-3-2\ell}
x_{12}^{-4} x_{11\p} x_{22\p}  .$$

We substitute the weight $\ell$ or the scaling helicity $h$ 
for   quarks as    $\ell= -1, h= -\frac{1}{2}$ and for 
 gluons as $\ell= -\frac{3}{2}, h= -1 $. 

In both the kernels for gluons and quarks  of parallel helicities 
only one value of the extra Yangian
representation parameter $v$ contributes, namely
$ v = 0 $.

In the kernel of  anti-parallel helicity quarks  
two values of $v$ contribute,  
$ v= 0, -1 $.
In the kernel of  anti-parallel helicity gluons 
three values of $v$ contribute, 
$ v= 0, -1 ,-2 $.

In the case of quarks 
$h=-\half$ the   contribution at  $v= -2$ is singular.  
Physically such a contribution occurs in the flavor singlet
channel only and it is proportional to $\delta(x_{1\p} - x_{2\p})$.
We note that the comparison also works in the case of gluon-quark
interaction kernels with $h_1 = -1, h_2= -\half$ and
$v\to \half, -\half,-\frac32 $. 

The leading order renormalization in the $SL(2)$ sector of $\mathcal{N} = 4$
super Yang Mills corresponds to the kernel (\ref{kernelRv}) 
at $h_1=h_2=0 $ and $v\to 0$.

In the case of ordinary parton distribution kernels where the momentum
transfer vanishes
we have the result (\ref{4corrk}) of the symmetric correlator.
In order to obtain the kernels in one-dimensional momenta 
we have to remove the factor carrying the
scaling weights and to do the same substitution of
parameters as leading to $\mathrm{R}_{12}(v)$ in positions.
First we have to do the substitution (\ref{usub1}),
$$ \e = u_2-u_3 \to u_4-u_3- 2\ell_2 =
 v_4-v_3 + \ell_3 + \ell_4 +2. $$
In the last step we have used our convention about the relation between the
notations $u$ and $v$, $u_i = v_i + \ell_i$, and the relation of the weights in
the considered 4-point correlator, $2\ell_2 + 2\ell_4 +2=0$. 
Finally we change to helicities according to $2\ell_i +1 = 2 h_i$ to obtain
the substitution rule of $\e = u_2-u_3$ by $v= v_4-v_3$ after relabeling 
the points $1 2 3 4 \to 1\p 2\p 1 2$ as   
$$ \e \to v+h_1 + h_2 +1. $$
This results in
\be \label{kernelk}
  (k_1 k_{1\p})^{-h_1-h_2 +v} (k_{1\p}-k_{1})^{-1-2v}. \ee
We use the momentum fraction $z= \frac{k_1}{k_{1\p}}$ and  consider the
same cases as above.  
The case of parallel helicity  appears in the extraordinary situations of 
quark structure functions of odd
chirality  and of  the photon structure function $F_3^{\gamma}$.
The kernels are \cite{XAM,SC} proportional to
$$ \frac{z}{(1-z)} , \ \ \ \ \frac{z^2}{(1-z)} $$
and they result as expected from (\ref{kernelk}) at $v=0$
and for $h_1 = h_2 = - \half$ and $h_1=h_2=-1$, respectively.

The  ordinary parton distributions have a scale dependence governed by
kernels composed as sums of contributions from (\ref{kernelk})
at $v=0, -1,-2$ for gluons and $v=0, -1$ for quarks. The coefficients
in the sums do not follow from the considered symmetry.
We find in comparison with the formulation in \cite{Levrep} 
for gluons ($h_1=h_2=-1$)
$$ z w_{+1}^{+1} = \frac{1+z^4}{(1-z)}= 2 \cdot \frac{z^2}{(1-z)} + 4 \cdot z (1-z) +
(1-z)^3, $$and for quarks ( $h_1 = h_2 = - \half$)
$$ w_{+\half}^{+\half} = \frac{1+z^2}{(1-z)} = 2 \cdot \frac{z}{(1-z)} +
(1-z).
$$   
$w_{+1}^{+1}$ and $w_{+\half}^{+\half}$ describe the probability distribution in
momentum fractions of the
 parton splitting of gluons or quarks with no flip of spin.
$w_{+ 1}^{-1}$ describes the gluon splitting with spin flip,
$$ z w_{+ 1}^{-1} = (1-z)^3. $$ 
It corresponds to our expression (\ref{kernelk}) at $h_1=h_2 = -1$ and $v=
-2$. A quark splitting with flip cannot occur in the flavor non-singlet
 channel at leading order,  $ w_{+\half}^{-\half} = 0 $. This corresponds to
 our statement that there is no $v=-2$ contribution in the quark case.

The scale evolution of the unpolarized parton distributions are determined
by the sum of the splitting kernels with and without spin flip, 
$w_{+1}^{+1}+w_{+1}^{-1}$, $ w_{+\half}^{+\half}+ w_{+\half}^{-\half}
=w_{+\half}^{+\half} $.  
  The scale evolution of the polarized (helicity asymmetry) parton distributions
is determined by the differences
$ w_{+1}^{+1} - w_{+ 1}^{-1} $ for gluons and 
$ w_{+\half}^{+\half}- w_{+ \half}^{-\half}= w_{+\half}^{+\half} $ for quarks (flavor non-singlet).
They correspond to our expression (\ref{kernelk}) at $v= 0, -1$
with the corresponding values of $h_1= h_2$.
 
In the expressions written above the regularization at $z=1, z=0$ is not
explicit. Recall that our definition of the integral operators assumes
integration contours which allow integration by parts (denoted as conjugation $^T$)
without boundary terms. Here the integration in $z$ over the segment
$(0,1)$ is to be understood as the result of the contraction of a closed
contour. This results in the conventional  formulation for parton splitting
kernels in terms of distributions like $\frac{1}{[1-z]_+}$.

\section{Spectra of symmetric integral operators} \label{sec:6}
\setcounter{equation}{0}

We consider operators defined in Sect. \ref{sec:5.1} by (\ref{intR}).
We choose as kernel a correlator with $2 M$ points $ 1\p, \dots, M\p, 1, \dots,
M$ with the weight balance 
$$ 2\ell_i + 2\ell_{i}\p + 2 = 0,\quad i= 1,\dots, M. $$
 This  implies 
$$  2\ell_i = -2 - 2\ell_{i}\p=  2 \tilde \ell_{i}. $$
In this way the weights of the resulting function $\mathrm{R} \psi$
are the same as the ones of $ \psi$.
This ensures the consistency of the eigenvalue problem on functions
depending homogeneously on $M$
two-dimensional points with the weights $2\tilde \ell_1,\dots,  2\tilde \ell_M $,
\be \label{ew}
\mathrm{R} \psi = \lambda   \psi. \ee

The Yangian symmetry of the operator induced by the symmetry of its kernel
has strong implications on the spectrum. In the case of the two-point
operator obeying the standard $RLL$ relation 
 the calculation of the spectrum is well known (cf. \cite{DKK00}). We recall
it for comparison with the generalized case. We describe the solution of the
spectral problem in the case of the symmetric 3-point operator. The outlined
scheme can be generalized to more points. 
For applications it is important to have  algorithms for calculating the
complete spectrum exactly.

\subsection{The spectrum of the YB operator}

The integral operator obtained from the kernel $\Phi^{--++}$ (\ref{kernel})
represents a Yang-Baxter operator intertwining the  representations
$2\ell_1$ and $2\ell_2$ according to (\ref{YBnormal}).  In this YB relation
we change to the ratio coordinate form (\ref{RLLv}) and   
write  the  $L$ matrix  in terms of $s\ell_2$
generators $S^a$ as (\ref{Sa}).
Let $\psi(x_1,x_2)$ obey the eigenvalue relation (the specification of
(\ref{ew})),
$$ \mathrm{R}_{21} \psi(x_1,x_2) = \lambda \psi(x_1,x_2). $$
 The YB relation implies
\be \label{Qv} 
\mathrm{R}_{21} L_{2}(v) L_{1}(v-\delta) \psi(x_1,x_2) = \lambda  
 L_{1}(v-\delta) L_2(v)  \psi(x_1,x_2).
\ee
We  abbreviate the relation as 
$ \mathrm{R} \mathcal{L} \psi= \lambda \mathcal{R} \psi $ 
and expand in powers of $v$. At $v^2$ we have a trivial relation, at $v^1$
we find:  
If $\psi$ is an eigenfunction then it is also $(S_1^a + S_2^a) \psi$ 
with the same eigenvalue, i.e. we have $s\ell_2$ irreducible representation subspaces  
of degeneracy.
The conditions at $v^0$ are related by this global symmetry. Therefore it is
sufficient to analyze the condition related to one matrix element, e.g.
$12$. 
 $$
\mathcal{R}_{12}= (S_1^0 -\delta) S_2^- - S_1^-S_2^0, 
\qquad \mathcal{L}_{12}= S_2^0 S_1^- - S_2^-(S_1^0 + \delta). $$
We consider the function
$$\psi_n^{(0)} = (x_1-x_2)^n $$ 
being a lowest weight vector
$$ (S_1^- +S_2^-) \psi_n^{(0)} = 0, \quad (S_1^0 +S_2^0) \psi_n^{(0)} = - \mu_n
\psi_n^{(0)}, \quad \mu_n = -\ell_1- \ell_2 +n. $$
We calculate  
$$ \mathcal{R}_{12} \psi_n^{(0)} =  
 \half \mu_n \mathrm{S}^- \psi_n^{(0)} + \half (1+\delta) \mathrm{S}^-
\psi_n^{(0)}, \quad 
 \mathcal{L}_{12} \psi_n^{(0)} = 
-\half \mu_n \mathrm{S}^- \psi_n^{(0)} - \half (1-\delta) \mathrm{S}^-
\psi_n^{(0)}, $$
where $\mathrm{S}^a= S_1^a - S_2^a$.
We obtain by (\ref{Qv})
$$ \lambda_n [\mu_n + 1 + \delta] \, \mathrm{S}^- \psi_n^{(0)} =  
[ - \mu_n + \delta -1] \, \mathrm{R} \, \mathrm{S}^- \psi_n^{(0)}. $$
We have marked the eigenvalue of the representation generated from the
lowest weight vector $\psi_n^{(0)}$ by the index $n$. 
We find that also $\mathrm{S} ^- \psi_n^{(0)}$ is an eigenfunction of
$\mathrm{R}_{21}$
with another eigenvalue,
$$ \mathrm{R}_{21} \ \mathrm{S} ^- \psi_n^{(0)} = \lambda_{n-1} \mathrm{S} ^-
\psi_n^{(0)}. $$
By comparison we obtain a recurrence relation for $\lambda_n$.
It is easy to see that the constant function is an eigenfunction and moreover
a lowest weight vector. Thus as the result of the analysis 
we  find all eigenfunctions and eigenvalues.

\subsection{The spectrum of a generalized YB operator}

We turn to the   6-point correlator $\Phi_6$ (\ref{6corr1})
obeying (\ref{YSC}) with the monodromy (\ref{T6}).

We relabel the points as $1,2,3,4,5,6 \to 1\p,2\p, 3\p,1,2,3$ and
consider the integral operator acting as (\ref{defQ123}).
We have shown above that this operator obeys
the generalized YB relation (\ref{genYB3}),  
$$
\hat  \R_{321}(\eps_4,\eps_5) L^+_3(u) L^+_2(\e_5+u) L^+_1(\e_4 + \e_5 +u)
=
L^+_1(\e_4 + \e_5 +u) L^+_2(\e_5+u) L^+_3(u) \hat \R_{321}(\eps_4,\eps_5). 
$$

We transform to the ratio coordinate form, where the $L$ operators are
substituted as (\ref{Sa}),
abbreviate the relations as  
$$ \hat  \R_{321} \mathcal{L}(u) = \mathcal{R}(u) \hat  \R_{321}, $$
 and expand
$$\mathcal{L}(u)= I u^3 + u^2 \mathcal{L}^{[2]} + u \mathcal{L}^{[1]}
+\mathcal{L}^{[0]}, \qquad \mathcal{R}(u)= I u^3 + u^2 \mathcal{R}^{[2]} + u \mathcal{R}^{[1]}
+\mathcal{R}^{[0]} .$$

The expansion in powers of $u$ results at $u^2$ in the
global symmetry condition
$$ [(S_1^a + S_2^a+S_3^a), \hat  \R_{321} ] = 0.  $$
It implies that the eigenvectors of $\hat  \R_{321}$ build
representations with the lowest weight vectors obeying
\be
\label{lw3} 
(S_1^- + S_2^- + S_3^-) \psi^{(0)}_{n} = 0, \ \ \ 
(S_1^0 + S_2^0 + S_3^0) \psi^{(0)}_{n} = [n -\ell_1 - \ell_2 -
\ell_3] \psi^{(0)}_{n}, \ee
and appearing explicitly as linear combinations at fixed $n=n_1+n_2$
\be \label{rin1}
 \psi^{(0),i}_{n} = \sum_{n_1 = 0}^n r(n)^{i}_{n_1} \psi^{(0)}_{n_1,n-n_1},
\ee 
of the basis lowest weight functions
\be  \label{psin1n2}
\psi^{(0)}_{n_1,n_2} = (x_1-x_2)^{n_1} (\frac{x_1+x_2}{2} - x_3)^{n_2}.
\ee
The problem reduces to find the particular linear combination obeying the 
eigenvalue relation 
\be \label{ewin}
 \hat  \R_{321} \psi^{(0),i}_{n} = \lambda_n^{(i)} \psi^{(0),i}_{n}. \ee
The degeneracy space of lowest weight vectors at given $n$ is $n+1$
dimensional.

Because of the global symmetry it is enough to consider one matrix element
of this relation; it is convenient to choose  the matrix element $12$.  
We shall calculate the expansion coefficients in 
\be \label{Rs}    
 \mathcal{R}^{[s]}_{12} \psi^{(0)}_{n_1,n-n_1} = \sum_k A^{[s]}_{+, k}(n-1,n_1)
\psi^{(0)}_{n_1+k,n-n_1-k-1} \ee
and similar in $\mathcal{L}^{[s]}_{12}$  with  $A^{[s]}_{-, k}(n-1,n_1)$,
 at $u^s$, $s=1,0$. 

 We substitute (\ref{rin1}) into (\ref{ewin})
and use the result of the action of ${\mathcal R}_{12}$ and ${\mathcal L}_{12}$.

$$ \hat  \R_{321} \sum_k A_{-,k} r(n)^i_ {n_1}
\psi^{(0)}_{n_1+k,n-1-n_1 -k} = 
\sum_k A_{+,k} r(n)^i_ {n_1} \lambda^{(i)}_n                     
\psi^{(0)}_{n_1+k,n-1-n_1 -k}. $$
The action of $\hat  \R_{321}$ has been calculated on r.h.s. by the eigenvalue relation
(\ref{ewin}). For doing this also on l.h.s. we express on both sides
$\psi^{(0)}_{n_1+k,n-1-n_1 -k}$ in the eigenfunction basis by the inversion
of the linear transformation (\ref{rin1}).
$$ \psi^{(0)}_{n_1,n-n_1} = \sum_i (  r^{-1}(n))^{n_1}_i 
\psi^{(0),i}_{n}.  $$
The coefficients at the eigenfunctions $\psi^{(0),j}_{n-1}$ obey the matrix
 equation (with the matrix elements labelled by $i= 0, \dots, n$ and
$j=0,1,\dots,n-1$)
$$ \sum_{k, n_1} A_{-,k} ( r(n))^i_{n_1} 
( r^{-1}(n-1))^{n_1+k}_j \lambda_{n-1}^{(j)} = 
\sum_{k, n_1} A_{+,k} ( r(n))^i_{n_1}   
( r^{-1}(n-1))^{n_1+k}_j \lambda_{n}^{(i)}.
$$

%%%%%%%%%%%%%%%%%%%%%%%%%%%%%%%%%%%%%%%%%%%%%%%%%%%%%%%%%%
We multiply by $( r^{-1}(n))^p_i \,  (r(n-1))^j_q $ and sum over
$ i=0, \dots,n$ and $ j= 0, \dots,n-1$.
\begin{multline} \sum_{k,j,n_1} \delta^p_{n_1}  \left (  (r^{-1}(n-1))^{n_1 + k}_j \lambda^{(j)}_{n-1} 
(r(n-1))^j_q \right )  A_{-,k}(n-1, n_1)
\\
= \sum_{k,i,n_1}
A_{+,k}(n-1, n_1)
\left (  (r^{-1}(n))^{p}_i \lambda^{(i)}_{n} 
(r (n))^i_{n_1} \right ) \delta^{n_1+k}_q.
\end{multline}
We put the eigenvalues $\lambda^{(i)}_n$ at fixed $n$ into the diagonal of a $n+1$
dimensional matrix and introduce the matrix obtained by similarity
transformation with the matrix $ r(n)$,
\be \label{Lambda}
 \Lambda(n) =  r^{-1}(n) \cdot \mathop{\mathrm{diag}}(\lambda_n)  \cdot r(n). \ee
Further, we introduce the rectangular $(n+1) \times n$ dimensional matrices
$\hat A_{\pm}$ with the elements
\be \label{Amatrix}
 (\hat A_{\pm})^p_q = \sum_k \delta^p_{q-k} A_{\pm, k}(n-1,
q-k). \ee
Then the relation reads
\be \label{iter3}
(\hat A_{-})^p_{q_1}  (\Lambda(n-1))^{q_1}_q =  (\Lambda(n))^{p}_{p_1}
(\hat A_{+})^{p_1}_q
\ee
$ p, p_1 = 0,\dots,n$ and $q,q_1 = 0, \dots, n-1$. 

We have $n(n+1)$ conditions at level $n$. These conditions with both
substitutions $\hat A_{\pm} \to \hat A^{[s]}_{\pm}$, $s=0,1$, (\ref{Rs}) 
are needed to find the $(n+1) \times (n+1)$ matrix elements of $\Lambda(n)$.

The matrices $\hat A^{[s]}_{\pm}$ encode the details of the Yangian symmetry.
The matrix $ r (n)$ transforms the basis $\psi^{(0)}_{n_1, n-n_1}$ to the
eigenfunction basis $\psi^{(0), i}_n $. 

The iterative condition (\ref{iter3}) allows to
calculate the matrices $\Lambda(n)$ with $\Lambda(0) = \lambda_0 $ as input.
The diagonalization
of this matrix then results in the eigenfunctions and eigenvalues in the
subspace of the irreducible representations in the threefold tensor product with
weight $2\ell_n = 2\ell_1+2\ell_2+2\ell_3 -n,   $ at the level $n=n_1+n_2$.

By calculating the action of the
operators $\mathcal{R}, \mathcal{L}$ appearing in the relation (\ref{genYB3})
at $u^1$ and $u^0$  on the basis (\ref{psin1n2}) 
we shall obtain in Appendix \ref{app:D} the explicit forms of the matrices
$\hat A^{[1]}_{\pm}$ and $\hat A^{[0]}_{\pm}$.  

As the result we find that they have
the following non-vanishing elements, at  $u^1$ 
\begin{align} & (\hat A^{[1]}_{\pm})^q_q = - (n-q) \Big\{\big( \half \e_4 +\e_5\big)\pm \big(n-1 -\half
\ell_1 - \frac{3}{2} \ell_2 - \ell_3 \big)  \Big\},  \notag\\
 & (\hat A^{[1]}_{\pm})^{q+1}_q = - (q+1) \left\{\e_4 \pm (q-\ell_1 -\ell_2)
\right\}, \notag\\
 \label{A1}
 & (\hat A^{[1]}_{\pm})^{q-1}_q  = \pm \frac{1}{4} (n-q+1) (n-q), 
\end{align}
and at   $u^0$
\begin{align} & (\hat A^{[0]}_{\pm})^q_q = (n-q)\Big\{ \Big[-\ell_1 \ell_2 -
\half q(q-1-\ell_1 - 3 \ell_2) +\half (\ell_1 - \ell_2)(n-q-1-\ell_3)\Big] \notag\\
&\hspace{3cm} \mp \Big[ \e_5(n-1-\ell_1-\ell_2-\ell_3)+\half\e_4
(n-1-\ell_3-2\ell_2) \Big]
- \e_5 (\e_4+\e_5)\Big\}, \notag\\
&(\hat A^{[0]}_{\pm})^{q+1}_q = (q+1) \Bigl\{ (q-\ell_1-\ell_2) (n-q-1- \ell_3) \pm
\e_4
(n-q-1 - \ell_3) \Bigr\}, \notag\\
& (\hat A^{[0]}_{\pm})^{q-1}_q = \frac{1}{4} (n-q+1) (n-q)
\Bigl\{ q-1-n -\ell_1 +3\ell_2 +\ell_3 \pm\e_4 \Bigr\},  \notag\\
 \label{A0}
& (\hat A^{[0]}_{\pm})^{q-2}_q = \frac{1}{8} (n-q+2) (n-q+1) (n-q).
\end{align}

A convenient algorithm is based on  the observation that the matrices
$\hat A_{\pm}$ are near diagonal. A quadratic diagonal matrix
$Z(n)$ can be found, 
$$ Z(n)_{q q} = \ell_3 -n+q+1, \quad q=0,1,\dots,n-1, $$
such that 
$\hat A^{[1]}_+ Z(n) + \hat A^{[0]}_+ = \hat A\p_+$ is upper triangular and has 
zero elements on the last row. The two conditions (\ref{iter3}) with
$\hat A^{[1]}_{\pm}$ and $\hat A^{[0]}_{\pm}$ can be combined in such
a way that  $\hat A\p_+$ appears on r.h.s. This linear combination
of the conditions allows to
calculate easily the matrix elements $\Lambda(n)^p_{p_1}$ with
$p= 0,1,\dots,n$ and $p_1= 0,1,\dots,n-1$. After this the matrix $\Lambda(n)$
with unknown elements on the last column only is substituted in one of the
initial conditions, e.g. in the one involving $\hat A^{[1]}_{\pm}$ (\ref{A1}),
to calculate these remaining elements.

\section{Discussion} \label{sec:7}
\setcounter{equation}{0}

The approach of Yangian symmetric correlators provides a simple way to
treat representations of the Yangian algebra in particular in the case of
Jordan-Schwinger type realizations of $s\ell_n$. The considered case of
$s\ell_2$ has the advantage that several explicit forms of the relevant
Yang-Baxter operators are known. This provides flexibility in the
calculations and allows formulations beyond those parallel to the methods
developed for scattering amplitudes. 
It is convenient
to use the forms of the $L$ and $R$ operators both in homogeneous  and in 
ratio coordinates. If the ratio coordinate $x$ describes position  
by transformation to the
helicity form one obtains the corresponding momentum dependence ($k= 
\lambda \bar \lambda $).

The monodromy matrix includes in the expansion in the spectral parameter $u$
($u_i = u+\delta_i$) the generators of the considered 
Yangian algebra representation. The latter is characterized by the conformal
weights $2\ell_i$ and the shifts $\delta_i$, which are the differences of
the spectral parameters in  $L_i(u_i)$, $u_{i+1} - u_i $. The YSC defined by
the monodromy depends on these representation parameters.
The described $R$ operator construction results in the weights calculated as
spectral parameter differences as well, moreover the sum  of all weights is fixed by
the elementary correlators from which the construction starts. This means
that we deal with particular representations where the
representation parameters are not all independent. We have encountered
examples with more than one relation constraining the weights. In such cases not all
spectral parameter differences or shifts are fixed by the weights, 
some differences appear as extra parameters. These extra parameters together with the
independent weights characterize the representation and the related YSC. 
In the considered example of a 6-point YSC we have 2 extra
parameters. This example can be generalized to $2M$-point YSC with
$M-1$ extra parameters.

The Yangian symmetry  condition (\ref{YSC}) in terms of the monodromy matrix
allows generalizations, where the $L$ matrices in the latter are replaced
by the ones with trigonometric or elliptic deformations or by other
forms related to the standard one by Drinfeld twists. It  would be
interesting to use such opportunities in applications beyond the well studied
integrable models.

The construction of YSC starting from elementary correlators relies on $R$
operators obeying Yang-Baxter relations with the product of two $L$
operators. These relations imply simple rules for the interchange of the $R$
operators with the product of $L$ operators in the monodromy. 

Actually the
$L$ operators, which have been written in several forms, are the source of
the rich symmetry structures. The simplicity of the $L$ operator in the case
of $s\ell_2$ allows to write  explicitly several kinds and representations
of $R$ operators.  

We have considered symmetric operators with YSC as kernels. The Yangian
symmetry of the kernel implies the symmetry of the operator. This follows
from the $L$ operators by their relations of inversion and transposition.
 
Particular 4-point YSC result in the different kinds of YB operators. 
The considered 6-point YSC results in a generalized YB operator and the
mentioned generalization of the latter to $2M$ points lead to generalized
YB operators where the counterpart of (\ref{genYB3}) involves the products of
$M$ factors of $L$ matrices on both sides.  The spectral parameter arguments
of the generalized YB operators are related to the extra parameters it their
kernels.

As a tool of treating Yangian symmetry the YSC approach is expected to be
useful in physical applications. In this respect it is important that we
have seen explicitly how the symmetry of the correlators implies the
symmetry of the operators constructed out of them and how the high symmetry
allows to solve exactly the spectral problem.   
We have formulated this in the non-trivial case of a 3-point generalized YB
operators constructed from a particular YSC. The outlined procedure
generalizes to more points with increasing  complexity.

As an example of applications we have shown the direct relation of particular
4-point YSC to 
the kernels of the scale evolution of (generalized) parton distributions
in leading order QCD. They are the ones
appearing as kernels of the YB operator $\R_{12}(v)$ intertwining the
tensor product of the representations with the weights $2\ell_1, 2\ell_2$. 
We have identified
the values of the weights and of the parameter $v$ appropriate for 
different cases of parton distributions, of gluons or quarks and of
different polarization configurations. In QCD  the parton
kernels are sums of contributions of 4-point YSC with different values
of the additional parameter.

These 4-point kernels describe simultaneously 
the scale dependence of a class of 
composite operators, actually the ones composed  of two quasi-partonic
field operators of gluons or quarks.
 The considered 6-point YSC is to be applied
to the scale dependence of composite operators out of 3 quark or gluon
field operators. 
 
In this way we have a reformulation in terms of YSC of the known relation 
of the scale
dependence of composite operators to the dynamics of quantum spin chains. 

The successful integrable spin chain treatment 
of all composite operators (with an arbitrary number of
fields) of $\mathcal{N}=4$ super Yang-Mills is related to the fact that 
here in the renormalization of analogous composite operators 
only the contribution  of one value of the additional
parameter,  $v\to 0$, appears.  Whereas in QCD in general more than one Yangian
representation contributes,  here one encounters only one contribution.

It would be interesting to consider the YSC based on the  
$s\ell(4|4)$ superalgebra  for a unified treatment of the quantum
integrability properties of $\mathcal{N}=4$ super Yang-Mills, i.e.
 to have on one hand  the known way to the scattering
amplitudes and on the other hand  an alternative way to the
treatment of the operator product renormalization.

\section*{Acknowledgments}

The authors are grateful to A.P. Isaev and \v{C}. Burd\'{i}k for discussions and
support. 

Our work has been supported by JINR Dubna via a Heisenberg-Landau grant.
The work of J.F. has been  supported by the Grant Agency of the Czech Technical
University in Prague, grant No. $SGS15/215/OHK4/3T/14$ and by the Grant of
the Plenipotentiary of the Czech republic at JINR, Dubna. 
% He thanks Leipzig University for hospitality.

%\vspace{1cm}

\appendix

\section{YB operator $R^{++}$ as contour integral} \label{app:A}

We want to obey the $RLL$ relation with the ansatz
$$ R_{12}^{++}(u) = \int \mathrm{d}c\, \phi(c)  e^{-c (\mathbf{x}_1 \cdot \mathbf{p}_2)
}. $$
We calculate the action of the shift operator on the product of $L$
matrices:
$$ e^{-c (\mathbf{x}_1 \cdot \mathbf{p}_2)}  \mathbf{p}_1 = 
(\mathbf{p}_1 + c \mathbf{p}_2 ) e^{-c (\mathbf{x}_1 \cdot \mathbf{p}_2)},
\qquad e^{-c (\mathbf{x}_1 \cdot \mathbf{p}_2)} \mathbf{x}_2 =
(\mathbf{x}_2 - c \mathbf{x}_1) \ e^{-c (\mathbf{x}_1 \cdot \mathbf{p}_2)},
$$ \vspace{-20pt}
\begin{multline*} e^{-c (\mathbf{x}_1 \cdot \mathbf{p}_2) } L_1^+ (u) L_2^+ (v) = 
\big(L_1^+ (u) + c \mathbf{p}_2 \otimes \mathbf{x}_1 \big)
\big(L_2^+ (v) - c \mathbf{p}_2 \otimes \mathbf{x}_1 \big)
e^{-c (\mathbf{x}_1 \cdot \mathbf{p}_2)}
\\
=\Big( L_1^+ (v) L_2^+ (u) + [L_2^+ (0) - L_1^+ (0) ] [u-v + c
(\mathbf{x}_1 \cdot \mathbf{p}_2) ]
- c [ u-v -1 + c (\mathbf{x}_1 \cdot \mathbf{p}_2) ]  \Big)
e^{-c (\mathbf{x}_1 \cdot \mathbf{p}_2)}.
\end{multline*}
The condition that the second term vanishes implies a differential equation
on $\phi(c) $, because
$$ \int \mathrm{d}c\, \phi(c)  \cdot c\cdot (\mathbf{x}_1 \cdot \mathbf{p}_2)
e^{-c (\mathbf{x}_1 \cdot \mathbf{p}_2)} = 
- \int \mathrm{d}c\, \phi(c) \cdot  c \cdot \dd_c\,
e^{-c (\mathbf{x}_1 \cdot \mathbf{p}_2)} = 
\int \mathrm{d}c\, \dd_c  (c \phi(c) )  \, e^{-c (\mathbf{x}_1 \cdot \mathbf{p}_2)}. $$ 
In the last step we have assumed that the integration by parts is done
without boundary terms.
Thus the condition on $\phi$ is
$$ \dd_c ( c \phi(c)) +  (u-v) \phi(c) = 0. $$
It is solved by
$$ \phi(c) = \frac{1}{c^{1 + u-v} }. $$
The condition of vanishing of the third term can be written as
$$ 0 = \dd_c( c^2 \phi(c)) + c (u-v-1)\phi(c) = 
c [ \dd_c( c \phi(c)) + (u-v)\phi(c) ].
$$
We see that it does not imply a further condition on $\phi(c)$.
Thus we have proved that the form used in calculation obeys
the Yang-Baxter relation provided the simple rule of integration by parts.
If the latter rule is different then the indicated procedure leads to the
appropriate modification. 

\section{ Calculation of the 6-point correlator}
\label{app:B}

We provide details of calculations leading to the correlator \eqref{6corr1}. 
The $\delta$-functions can be used to fix six out of the nine integration variables as the
solution of ($i=1,2,3$)
$$ x_{i, 1} - c_{i4} x_{4, 1} -c_{i 5} x_{5,1} -c_{i 6} x_{6,1} = 0, \ \ \
x_{i, 2} - c_{i4} x_{4, 22} -c_{i 5} x_{5,2} -c_{i 6} x_{6,2} = 0. 
$$
We let $c_{i6}$ free and find $c_{i4}, c_{i 5}$ and the Jacobi factor
 $\langle 45 \rangle^{-3} $,
$$
c_{i4}^{(0)} = \frac{\langle i5 \rangle -c_{i6} \langle 65\rangle}{\langle 45 \rangle}, \ \ \
c_{i5}^{(0)} = \frac{\langle i4 \rangle -c_{i6} \langle 64\rangle}{\langle 54 \rangle}, $$
$$  
\prod_{i=1}^3 \delta^{(2)} \big(\mathbf{x}_i -  \sum_{j=4}^6 c_{ij}
\mathbf{x}_j \big) = \langle 45\rangle^{-3} \prod_{i=1}^3 \delta\big(c_{i4} - c_{i4}^{(0)}\big) 
\delta\big(c_{i5} - c_{i5}^{(0)}\big).   $$
The minors result in
$$ (2 3 4) \to c_{14}^{(0)} = \frac{\langle 15\rangle - c_{16} \langle 65\rangle}{\langle 45\rangle}, $$
$$
(3 4 5) \to 
\left \vert \begin{matrix}
c_{14}^{(0)}& c_{15}^{(0)}  \cr
c_{24}^{(0)}& c_{25}^{(0)}  
\end{matrix} \right \vert 
 =  \langle 45\rangle^{-2} \ 
\left \vert \begin{matrix}
\langle 15\rangle- c_{16}\langle 65\rangle   & \ \langle 14 \rangle - c_{16} \langle 64\rangle   \cr
\langle 25\rangle -c_{26} \langle 65\rangle  & \ \langle 24\rangle - c_{2 6} \langle 64\rangle   
\end{matrix} \right \vert. 
$$
We use relations of the type 
$$
\left \vert \begin{matrix}
\langle 15\rangle& \langle 14\rangle  \cr
\langle 25 \rangle & \langle 24\rangle     
\end{matrix} \right \vert
= \langle 12\rangle \,  \langle 54\rangle
$$ and obtain
$$ (3 4 5) \to \langle 45\rangle^{-1} \left( \langle 12\rangle -c_{16} \langle 62\rangle + c_{26} \langle 61\rangle \right), $$
$$ (4 5 6 )\to \langle 45\rangle^{-2} 
\left \vert \begin{matrix}
c_{14}^{(0)}& c_{15}^{(0)} & c_{16}  \cr
c_{24}^{(0)}& c_{25}^{(0)} & c_{26} \cr
c_{34}^{(0)}& c_{35}^{(0)} & c_{36} 
\end{matrix} \right \vert   
= 
\langle 45\rangle^{-1} \left ( \langle 12\rangle c_{36} + \langle 23\rangle c_{16} + \langle 31\rangle c_{26} \right ), $$
$$ (5 6 1) \to \langle 45\rangle^{-1} 
\left \vert \begin{matrix}
c_{25}^{(0)}& c_{26}  \cr
c_{35}^{(0)}& c_{36}
\end{matrix} \right \vert = 
 \langle 45\rangle^{-1} ( \langle 24\rangle c_{36} - \langle 34\rangle c_{26} ), $$
$$ (6 1 2) \to c_{36}. $$

\section{ q-deformed YB operator $\mathcal{S}_{q 12}$ }
\label{app:C}

We use the abbreviation 
$$\Phi(x,\lambda) = \frac{(xq^{1-\lambda};q^2)}{(xq^{1+\lambda};q^2)}$$ 
defined using the infinite product 
$(x;q):= \prod_{j=1}^\infty (1-xq^{j-1}).$
Because $\mathcal{S}_{q12}(w)$ involves the coordinates only, it commutes with 
the matrix factors $\hat V_i$ in the factorization \eqref{L+r}.  Commuting it through the matrix factors $\hat D_i$ creates extra matrix factors.
Commuting $\mathcal{S}_{q12}(v_1^{(2)}-v_2^{(1)})$ through 
$L_1(v_1^{(1)},v_1^{(2)})L_2(v_2^{(1)}, v_2^{(2)})$ 
we see that it commutes with the first factor of $L_1$ and the third factor of $L_2$. 
We get from the second and third factor of $L_1$ and the first and second factor of $L_2$
\begin{align*}
& \begin{pmatrix}
q^{-v_1^{(2)}+v_2^{(1)}} \frac{\Phi(qx_2/x_1,v_1^{(2)}-v_2^{(1)})}{\Phi(x_2/x_1,v_1^{(2)}-v_2^{(1)})} 
& 0 \\ 0 & q^{v_1^{(2)}-v_2^{(1)}} \frac{\Phi(q^{-1}x_2/x_1,v_1^{(2)}-v_2^{(1)})}{\Phi(x_2/x_1,v_1^{(2)}-v_2^{(1)})}
\end{pmatrix}
\begin{pmatrix}
q^{v_1^{(2)}} & x_1^{-1} \\ -q^{-v_1^{(2)}} & -x_1^{-1} 
\end{pmatrix}
\\
& \hspace{3cm} \times 
\begin{pmatrix}
1 & 1 \\ -x_2 q^{-v_2^{(1)}} & -x_2 q^{v_2^{(1)}}
\end{pmatrix}
\begin{pmatrix}
\frac{\Phi(x_2/x_1,v_1^{(2)}-v_2^{(1)})}{\Phi(q x_2/x_1,v_1^{(2)}-v_2^{(1)})}   & 0 
\\ 0 & \frac{\Phi(x_2/x_1,v_1^{(2)}-v_2^{(1)})}{\Phi(q^{-1}x_2/x_1,v_1^{(2)}-v_2^{(1)})}  
\end{pmatrix} 
\\
& = \begin{pmatrix}
q^{v_2^{(1)}} - \frac{x_2}{x_1}q^{-v_1^{(2)}} & q^{v_2^{(1)}} -\frac{x_2}{x_1} q^{v_1^{(2)}} \\
 \frac{x_2}{x_1}q^{-v_1^{(2)}}-q^{-v_2^{(1)}} &  \frac{x_2}{x_1} q^{v_1^{(2)}} -q^{-v_2^{(1)}}
\end{pmatrix}  
=\begin{pmatrix}
q^{v_2^{(1)}} & x_1^{-1} \\ -q^{-v_2^{(1)}} & -x_1^{-1} 
\end{pmatrix}
\begin{pmatrix}
1 & 1 \\ -x_2 q^{-v_1^{(2)}} & -x_2 q^{v_1^{(2)}}
\end{pmatrix}
\end{align*}
which are the third factor of $L_1$ and the first factor of $L_2$ on the right hand side of  (\ref{eq:RLL-rational}). 

The proof can be repeated also in terms of the homogeneous coordinates $x_{i,1},x_{i,2}$.
The $L^+$ operator (\ref{Lq+-}) can be reconstructed  from the one in the 
ratio coordinates (\ref{L+r}) by inserting $x_1 = \frac{x_{1,1}}{x_{1,2}} $ and $2\ell_1 \to x_{1,1}
\dd_{1,1}+ x_{1,2} \dd_{1,2}$ and in analogous way also the $L^-$ operator.
% by inserting $x_2=-\frac{x_{2,2}}{x_{2,1}}$ and $2\ell_2 \to x_{2,1} \dd_{2,1}+ x_{2,2} \dd_{2,2}$. 

To simplify formulas we use in the rest of this appendix the following notation: $x_1=x_{1,1}$, $x_2=x_{1,2}$, $y_1=x_{2,1}$, $y_2=x_{2,2}$. 
The derivatives appearing below are always multiplied by the respective variables and we 
adopt the following simplified notation: $x_1\partial_1=x_{1,1}\partial_{1,1}$, $x_2\partial_2=x_{1,2}\partial_{1,2}$, 
whereas $y_1\partial_1=x_{2,1}\partial_{2,1}$, $y_2\partial_2=x_{2,2}\partial_{2,2}$. The q-deformed R operator 
\begin{equation}
R(\lambda) = (x_1y_1)^\lambda \cdot \Phi(\mathcal{X},\lambda),
\end{equation}
where $\mathcal{X}=-\frac{x_2y_2}{x_1y_1}$, commutes with factors of the L operators as
$$
 R \begin{pmatrix}
1 &1 \\ \frac{x_1}{x_2} q^{-u -x_1\partial_1 -x_2\partial_2} &
\frac{x_1}{x_2} q^{u +x_1\partial_1 +x_2\partial_2}
\end{pmatrix} = \begin{pmatrix}
1 &1 \\ 
\frac{x_1}{x_2} q^{-u+\lambda -x_1\partial_1 -x_2\partial_2} &
\frac{x_1}{x_2} q^{u-\lambda +x_1\partial_1 +x_2\partial_2}   
\end{pmatrix} R, 
$$ $$
 R \begin{pmatrix}
q^{x_1\partial_1+1} & 0 \\ 0 & q^{-x_1\partial_1 -1}
\end{pmatrix} = 
\begin{pmatrix} 
q^{x_1\partial_1+1-\lambda}
\frac{\Phi(q\mathcal{X},\lambda)}{\Phi(\mathcal{X},\lambda)} & 0 \\
0 & q^{\lambda-x_1\partial_1 -1}
\frac{\Phi(q^{-1}\mathcal{X},\lambda)}{\Phi(\mathcal{X},\lambda)}
\end{pmatrix} R, $$ $$
 R \begin{pmatrix}
q^{y_2\partial_2+1} & 0 \\ 0 & q^{-y_2\partial_2 -1}
\end{pmatrix} = 
\begin{pmatrix} 
\frac{\Phi(\mathcal{X},\lambda)}{\Phi(q\mathcal{X},\lambda)}
q^{y_2\partial_2+1} & 0 \\
0 &
\frac{\Phi(\mathcal{X},\lambda)}{\Phi(q^{-1}\mathcal{X},\lambda)}q^{-y_2\partial_2
-1}
\end{pmatrix} R, $$ $$
 R \begin{pmatrix}
q^{v-y_1\partial_1-y_2\partial_2-2} & \frac{y_1}{y_2} \\
-q^{-v+y_1\partial_1+y_2\partial_2+2} & -\frac{y_1}{y_2}
\end{pmatrix} =
\begin{pmatrix}
q^{v+\lambda -y_1\partial_1-y_2\partial_2-2} & \frac{y_1}{y_2} \\
-q^{-v-\lambda+y_1\partial_1+y_2\partial_2+2} & -\frac{y_1}{y_2} 
\end{pmatrix} R.
$$
After commuting $R(\lambda)$ through $L^+(u)L^-(v)$ we obtain from the second and the third factor of $L^+(u)$ and the first and the second
factor of $L^-(v)$
\begin{align*}
& \begin{pmatrix}
q^{-\lambda} \frac{\Phi(q\mathcal{X},\lambda)}{\Phi(\mathcal{X},\lambda)} & 0 \\ 0 & q^{\lambda} \frac{\Phi(q^{-1}\mathcal{X},\lambda)}{\Phi(\mathcal{X},\lambda)}
\end{pmatrix} 
\begin{pmatrix}
q^{u-1} & -\frac{x_2}{x_1} \\
-q^{-u+1} & \frac{x_2}{x_1}
\end{pmatrix}  
\begin{pmatrix}
1 &1 \\ -\frac{y_2}{y_1} q^{-v+1 } & -\frac{y_2}{y_1} q^{v-1}
\end{pmatrix} 
\\
&\hspace{10cm} \times
\begin{pmatrix}
\frac{\Phi(\mathcal{X},\lambda)}{\Phi(q\mathcal{X},\lambda)}  & 0 \\ 0 & \frac{\Phi(\mathcal{X},\lambda)}{\Phi(q^{-1}\mathcal{X},\lambda)}
\end{pmatrix} 
\\
& = \begin{pmatrix}
q^{u-\lambda-1}+\mathcal{X} q^{-v-\lambda+1} & (q^{-\lambda+u-1}+\mathcal{X}
q^{-\lambda+v-1} )
\frac{\Phi(q\mathcal{X},\lambda)}{\Phi(q^{-1}\mathcal{X},\lambda)} \\
- (q^{\lambda-u+1}+\mathcal{X} q^{\lambda-v+1} )
 \frac{\Phi(q^{-1}\mathcal{X},\lambda)}{\Phi(q^{}\mathcal{X},\lambda)}&
- (q^{\lambda-u+1}+\mathcal{X} q^{\lambda+v-1} )
\end{pmatrix} 
\intertext{and setting $ \lambda=u-v$}
& = \begin{pmatrix}
q^{v-1} +\mathcal{X}q^{-u+1} & q^{v-1} + \mathcal{X} q^{u-1} \\
- (q^{-v+1} \mathcal{X} q^{-u+1}) & -(q^{-v+1} +\mathcal{X} q^{u-1})
\end{pmatrix} 
 = \begin{pmatrix}
q^{v-1} & -\frac{x_2}{x_1} \\
-q^{-v+1} & \frac{x_2}{x_1}  
\end{pmatrix}
\begin{pmatrix}
1 &1 \\ -\frac{y_2}{y_1} q^{-u+1 } & -\frac{y_2}{y_1} q^{u-1}
\end{pmatrix}
\end{align*}  
which proves (\ref{RqLL}).

\section{ Calculation of $\hat A_{\pm}$ }
\label{app:D}

At $u^1$ in the generalized YB relation (\ref{genYB3}) appears
\begin{multline*}
\hat{\mathbb{R}} \left ( \sigma S_3 \sigma S_2 + \e_5 \sigma S_3 + \sigma S_3 \sigma S_1
+
(\e_4+\e_5) \sigma S_3 + \sigma S_2 \sigma S_1 + \e_5 \sigma S_1 (\e_4+\e_5)
\sigma S_2 \right ) \\ 
= \left ( \sigma S_1 \sigma S_2 + (\e_4 +\e_5) \sigma S_2 + \e_5 \sigma S_1 
 + \sigma S_1 \sigma S_3+
(\e_4+\e_5) \sigma S_3 + \sigma S_2 \sigma S_3 + \e_5 \sigma S_3  \right )
\hat{\mathbb{R}}. \end{multline*}
We have abbreviated using Pauli matrices $\sigma S = \sigma ^a S^a $.
We consider the matrix element $12$,  
$$ \mathcal{R}_{12} = \mathcal{R}^{(1)}_{12} +\mathcal{R}^{(2)}_{12}, \qquad 
\mathcal{L}_{12} = \mathcal{L}^{(1)}_{12} +\mathcal{L}^{(2)}_{12},$$
$$ \mathcal{R}^{(2)}_{12}= -\mathcal{L}^{(2)}_{12}, \qquad
\mathcal{R}^{(1)}_{12}= \mathcal{L}^{(1)}_{12}, $$
$$  \mathcal{R}^{(1)}_{12} = \e_4 S_2^-+ (\e_4 + \e_5) S_3^-, $$
$$ \mathcal{R}^{(2)}_{12} = 
S_1^0 S_2^- - S_1^- S_2^0 + S_1^0 S_3^- - S_1^- S_3^0 +S_2^0 S_3^- - S_2^-
S_3^0.
$$ 
We have suppressed the superscript $[s]$ indicating the power of expansion in $u$.
 
We calculate the action of the operators involved on the basis 
$\psi^{(0)}_{n_1,n_2}  $. 
\begin{align}{\mathcal R}_{12}^{(1)} \psi^{(0)}_{n_1,n_2}  &  = 
- \e_4 \big(n_1 \psi^{(0)}_{n_1-1,n_2} +
\half n_2  \psi^{(0)}_{n_1,n_2-1} \big) - \e_5 n_2 \psi^{(0)}_{n_1,n_2-1}, 
\notag\\
 {\mathcal R}_{12}^{(2)} \psi^{(0)}_{n_1,n_2}  & =
- (S_1^0 + S_2^0) \big(n_1 \psi^{(0)}_{n_1-1,n_2} +\half n_2  \psi^{(0)}_{n_1,n_2-1} \big)
\notag \\
& \hspace{2.5cm} - (S_1^- + S_2^-) \big(-\ell_3 \psi^{(0)}_{n_1,n_2}- n_2 x_3
- \psi^{(0)}_{n_1,n_2-1}\big) - S_2^0 n_2 \psi^{(0)}_{n_1,n_2-1} 
\notag \\
&  = \psi^{(0)}_{n_1,n_2-1} \big( - \half n_2 (n_1 + n_2 -1 - \ell_1 - \ell_2) + n_2
(\ell_1 +\ell_2) \big) 
\notag\\
& \quad + \psi^{(0)}_{n_1-1,n_2} \big(- n_1 (n_1 + n_2 -1 - \ell_1 - \ell_2)\big) + \psi^{(0)}_{n_1-1,n_2-1} (-n_1 n_2 x_3 + n_1 n_2 x_2) \notag\\
& \qquad\qquad + \psi^{(0)}_{n_1,n_2-2} \big( - \half n_2 ( n_2-1) x_3 + n_2 (n_2-1) x_3 - n_2
(n_2-1) x_2 \big). 
\end{align}
We use 
$$ ( x_3-x_2) \psi^{(0)}_{n_1,n_2} = \half \psi^{(0)}_{n_1+1,n_2}
-\psi^{(0)}_{n_1,n_2+1} $$
and obtain
\begin{multline}
{\mathcal R}_{12}^{(2)} \psi^{(0)}_{n_1,n_2}  = 
-n_2 \big(n_1+n_2 -1 - \half \ell_1 -\frac{3}{2} \ell_2 - \ell_3 \big) \psi^{(0)}_{n_1,n_2-1}   \\
- n_1 (n_1 -1 - \ell_1 - \ell_2)
\psi^{(0)}_{n_1-1,n_2} + \frac{1}{4} n_2 (n_2-1) \psi^{(0)}_{n_1+1,n_2-2} . 
\end{multline}
The result of the operator action at $u^1$ can be formulated in terms of
the matrices $\hat A^{[1]}_{\pm}$ (\ref{Amatrix}) as written above
(\ref{A1}).

\vspace{0.5cm}

We consider also the $u^0$ contribution to (\ref{genYB3}).
With the abbreviation $\mathrm{R}\mathcal{L} = \mathcal{R} \mathrm{R}$
 we have
$$ \mathcal{L} = (\sigma S_3) (\sigma S_2)(\sigma S_1) + \e_5 (\sigma
S_3)(\sigma S_1) + (\e_4+\e_5) (\sigma S_3)(\sigma S_2) 
+ \e_5 (\e_4 + \e_5) (\sigma S_3),
$$
$$ \mathcal{R} = (\sigma S_1) (\sigma S_2)(\sigma S_3) + \e_5 (\sigma
S_1)(\sigma S_3) + (\e_4+\e_5) (\sigma S_2)(\sigma S_3)+ \e_5 (\e_4 + \e_5) (\sigma S_3).
$$
We shall use again the matrix element $12$ of the relation.
$$  \mathcal{R} =  \mathcal{R}^{(3)}+  \mathcal{R}^{(2)}+ \mathcal{R}^{(1)},
$$
$$ \mathcal{R}^{(3)}_{12} = S_1^0 S_2^0 S_3^- + S_1^- S_2^+ S_3^- - S_1^0
S_2^- S_3^0 + S_1^- S_2^0 S_3^0, $$
$$ 
\mathcal{R}^{(2)} = \e_5 (S_1^0 S_3^- - S_1^- S_3^0) + (\e_4 + e_5)
(S_2^0 S_3^- - S_2^- S_3^0), $$
 $$
\mathcal{R}^{(1)} = \e_5 (\e_4+e_5) S_3^-.
$$
We suppress the superscript $[s]$ indicating the expansion power in $u$.
Similar to the case $u^1$ we have 
$$ \mathcal{L}^{(3)} = \mathcal{R}^{(3)}, \quad \mathcal{L}^{(2)} =
- \mathcal{R}^{(2)}, \quad \mathcal{L}^{(1)} = \mathcal{R}^{(1)}. $$

We calculate the action of these operators on the basis
functions of lowest weight (\ref{psin1n2}).
We start with contributions to $\mathcal{R}^{(3)}_{12}$.
\begin{align*} & (S_1^0 S_2^0 + S_1^- S_2^+) \psi_{n_1,n_2}^{(0)} = \big\{ (x_1x_2 - x_2^2) \dd_1 \dd_2 - \ell_1 x_2 \dd_2 - \ell_2 x_1 \dd_1 +
2 \ell_2 x_2 \dd_1 +\ell_1 \ell_2 \big\}  \psi_{n_1,n_2}^{(0)}
\\
&\qquad  = x_2 (x_1-x_2) \big\{ - n_1 (n_1-1)  \psi_{n_1-2,n_2}^{(0)} + \frac{1}{4}
n_2(n_2-1)  \psi_{n_1,n_2-2}^{(0)} \big\}  +
\ell_1 \ell_2 \psi_{n_1,n_2}^{(0)}
\\
& \quad +  \ell_2 (2x_2 - x_1) \big\{ n_1  \psi_{n_1-1,n_2}^{(0)} + \half n_2
\psi_{n_1,n_2-1}^{(0)} \big\} -  \ell_1 x_2 \big\{-n_1  \psi_{n_1-1,n_2}^{(0)} + \half n_2
\psi_{n_1,n_2-1}^{(0)} \big\} . 
\end{align*}
We substitute
$$ x_2 = \big(\frac{x_1+x_2}{2} - x_3\big) - \half (x_1-x_2) + x_3 $$
and separate a contribution proportional to $x_3$.
\begin{align*}
& (S_1^0 S_2^0 + S_1^- S_2^+) \psi_{n_1,n_2}^{(0)}  
\\
& \,\, = x_3 \big\{ (-n_1 +1 +\ell_1 +\ell_2) n_1 \psi_{n_1-1,n_2}^{(0)} + \frac{1}{4}
n_2(n_2-1) \psi_{n_1+1,n_2-2}^{(0)} - (\ell_1-\ell_2) \half n_2
\psi_{n_1,n_2-1}^{(0)} \big\} 
\\
& \qquad +(-n_1 +1 +\ell_1 +\ell_2) n_1 \psi_{n_1-1,n_2+1}^{(0)} + \Big\{ \half (n_2-1) - \ell_2 + \half(\ell_1-\ell_2) \Big\} \half n_2 \psi_{n_1+1,n_2-1}^{(0)}  
\\
&  \, - \frac{1}{8}  n_2(n_2-1) \psi_{n_1+2,n_2-2}^{(0)}  +\big\{\ell_1 \ell_2 + \half n_1(n_1-1) - n_1( \half \ell_1 + \frac{3}{2} \ell_2) - 
\half n_2(\ell_1 - \ell_2)\big\} \psi_{n_1,n_2}^{(0)},
\end{align*}
\begin{align*} & S_3^- (S_1^0 S_2^0 + S_1^- S_2^+) \psi_{n_1,n_2}^{(0)} \\
&\quad= - x_3 \left\{ (-n_1 +1 +\ell_1 +\ell_2) n_1 n_2 \psi_{n_1-1,n_2-1}^{(0)} + \frac{1}{4}
n_2(n_2-1) (n_2 -2)\psi_{n_1+1,n_2-3}^{(0)} \right. 
\\
& \hspace{2.5cm} \left. - (\ell_1-\ell_2) \half n_2 (n_2-1)
\psi_{n_1,n_2-2}^{(0)} \right\} + (n_1 -1 -\ell_1 -\ell_2) n_1 n_2  \psi_{n_1-1,n_2}^{(0)}
\\
& \hspace{1.5cm} +\frac{1}{8}  n_2(n_2-1) (n_2-2)\psi_{n_1+2,n_2-3}^{(0)} - (n_2 -2 + \ell_1 - 3\ell_2 ) \frac{1}{4} n_2 (n_2-1) \psi_{n_1+1,n_2-2}^{(0)} 
\\
& \hspace{1.5cm} - \big\{2\ell_1 \ell_2 + n_1(n_1-1) - n_1(  \ell_1 + 3 \ell_2) -
 (n_2-1) (\ell_1 - \ell_2)\big\} \half n_2 \psi_{n_1,n_2-1}^{(0)},
\end{align*}
\begin{multline*} (S_1^0 S_2^- - S_1^- S_2^0 )  \psi_{n_1,n_2}^{(0)} 
 \\ = (- n_1 +1 + \ell_1 +\ell_2)] n_1  \psi_{n_1-1,n_2}^{(0)} + 
\frac{1}{4} n_2 (n_2-1)  \psi_{n_1+1,n_2-2}^{(0)} 
- (\ell_1-\ell_2) \half n_2  \psi_{n_1,n_2-1}^{(0)}, 
\end{multline*}
\begin{align*}
& S_3^0 (S_1^0 S_2^- - S_1^- S_2^0 )  \psi_{n_1,n_2}^{(0)} \\
& \quad = - x_3 \left\{ (- n_1 +1 + \ell_1 +\ell_2) n_1 n_2 \psi_{n_1-1,n_2-1}^{(0)}  \right.
\\
& \hspace{3cm} + \frac{1}{4} n_2 (n_2-1) (n_2-2)  \psi_{n_1+1,n_2-3}^{(0)} -
(\ell_1-\ell_2) \half n_2  (n_2-1) \psi_{n_1,n_2-2}^{(0)} \Big\} 
\\
& - \ell_3 \big\{ (- n_1 +1 + \ell_1 +\ell_2) n_1  \psi_{n_1-1,n_2}^{(0)} + 
\frac{1}{4} n_2 (n_2-1)  \psi_{n_1+1,n_2-2}^{(0)} -
(\ell_1-\ell_2) \half n_2  \psi_{n_1,n_2-1}^{(0)} \big\}.
\end{align*}

In the action of 
$  \mathcal{R}^{(3)}_{12} = (S_1^0 S_2^0  + S_1^- S_2^+) S_3^- - (S_1^0
S_2^-  + S_1^- S_2^0) S_3^0 $
the contributions proportional to $x_3$ cancel. It is important to find the
result containing lowest weight contributions only.
\begin{align}
&  \mathcal{R}^{(3)}_{12} \psi_{n_1,n_2}^{(0)} \notag\\
& \quad = n_1 ( n_1 -1 - \ell_1 -\ell_2)(n_2 - \ell_3) \psi_{n_1-1,n_2}^{(0)} +
\frac{1}{8} n_2 (n_2-1) (n_2-2)  \psi_{n_1+2,n_2-3}^{(0)} \notag
\\
&
\qquad+ n_2 \Big\{-\half(l_1-l_2)(l_3+1) -\ell_1\ell_2 -\half n_1(n_1-1-\ell_1-3\ell_2) + 
\half n_2(\ell_1-\ell_2) \Big\} \psi_{n_1,n_2-1}^{(0)} \notag
\\
& \hspace{5.7cm} -\frac{1}{4} n_2 (n_2-1) (n_2 + \ell_1 - 3\ell_2 - \ell_3) 
 \psi_{n_1+1,n_2-2}^{(0)}.  \label{R3}
\end{align}

The action of $\mathcal{R}^{(2)}$ is calculated easier.
\begin{multline}
\mathcal{R}_{12}^{(2)}  \psi_{n_1,n_2}^{(0)}=  \frac{1}{4} n_2
(n_2-1) \e_4 \psi_{n_1+1,n_2-2}^{(0)} + \e_4 n_1 (n_2-\ell_3) \psi_{n_1-1,n_2}^{(0)}  \\
 -n_2 \big\{ \e_5 (n_1+n_2-1 -\ell_1-\ell_2-\ell_3)+\half \e_4(n_1+n_2-1-\ell_3-2\ell_2) \big\}
 \psi_{n_1,n_2-1}^{(0)}.  \label{R32}
\end{multline}

We add also
\be
\mathcal{R}_{12}^{(1)}  \psi_{n_1,n_2}^{(0)}= 
 - \e_5 (\e_4+\e_5) n_2  \psi_{n_1,n_2-1}^{(0)} .\ee
According to our conventions the result can be written in terms of the
matrices $\hat A^{[0]}_{\pm}$ given above (\ref{A0}).

\end{document}